\documentclass[preprint2]{aastex}

\newcommand{\paperone}{Paper I}
\newcommand{\papertwo}{Paper II}
\newcommand{\paperthree}{Paper III}
\newcommand{\kms}{\mbox{km~s$^{-1}$}}
\newcommand{\whz}{\mbox{W~Hz$^{-1}$}}
\newcommand{\ergs}{\mbox{erg~s$^{-1}$}}
\newcommand{\ergsc}{\mbox{erg~s$^{-1}$~cm$^{-2}$}}
\newcommand{\ergscA}{\mbox{erg~s$^{-1}$~cm$^{-2}$\AA$^{-1}$}}
\newcommand{\ho}{\mbox{$H_{\, {\rm 0}}$}}
\newcommand{\qo}{\mbox{$q_{\, {\rm 0}}$}}
\newcommand{\lsun}{\mbox{$L_{\odot}$}}
\newcommand{\iras}{\emph{IRAS}}
\newcommand{\dbs}{DBS}
\newcommand{\iraf}{\emph{IRAF}}
\newcommand{\splot}{SPLOT}
\newcommand{\ngaussfit}{NGAUSSFIT}
\newcommand{\fris}{\mbox{FR~Is}}
\newcommand{\frii}{\mbox{FR~II}}
\newcommand{\cg}{CSS/GPS}
\newcommand{\nLnfir}{\mbox{${\nu}L_\nu$(60~$\mu$m)}}
\newcommand{\Snfir}{\mbox{$S_\nu$(60~$\mu$m)}}
\newcommand{\pfive}{\mbox{$P_{{\rm \, 4.8~ GHz}}$}}
\newcommand{\Snrad}{\mbox{$S_\nu$(4.8~ GHz)}}
\newcommand{\uexpr}{\mbox{$\log~[\Snfir/\Snrad]$}}
\newcommand{\oiii}{\mbox{[\ion{O}{3}]~$\lambda$~5007}}

\newcommand{\hbeta}{\mbox{H$\rm{\beta}$}}
\newcommand{\hdelta}{\mbox{H$\rm{\delta}$}}
\newcommand{\hgamma}{\mbox{H$\rm{\gamma}$}}
\newcommand{\hepsilon}{\mbox{H$\rm{\epsilon}$}}
\newcommand{\oiiia}{\mbox{[\ion{O}{3}]~$\lambda$5007}}
\newcommand{\oi}{\mbox{[\ion{O}{1}]~$\lambda$6300}}
\newcommand{\halpha}{\mbox{H$\rm{\alpha}$}}
\newcommand{\niia}{\mbox{[\ion{N}{2}]~$\lambda$6584}}
\newcommand{\niiab}{\mbox{[\ion{N}{2}]~$\lambda$6548, 6584}}
\newcommand{\siia}{\mbox{[\ion{S}{2}]~$\lambda$6717}}
\newcommand{\siib}{\mbox{[\ion{S}{2}]~$\lambda$6731}}
\newcommand{\lratioOIIIHb}{\mbox{$\log$~[\ion{O}{3}]/H$\beta$}}
\newcommand{\ratioNIIHa}{\mbox{[\ion{N}{2}]/H$\alpha$}}
\newcommand{\ratioSIIHa}{\mbox{[\ion{S}{2}]/H$\alpha$}}
\newcommand{\ratioOIHa}{\mbox{[\ion{O}{1}]/H$\alpha$}}

\normalsize

\begin{document}

\title{Radio-Excess \iras\ Galaxies: IV. Optical Spectroscopy}

\author{Catherine L.\ Buchanan\altaffilmark{*,1,2}, Peter J.\
McGregor\altaffilmark{1}, Geoffrey V.\ Bicknell\altaffilmark{1}, \and Michael
A.\ Dopita\altaffilmark{1}}

\altaffiltext{*}{Previously Catherine L. Drake}
\altaffiltext{1}{Research School of Astronomy and Astrophysics, The Australian National University, Cotter Rd, Weston, ACT 2611 Australia}
\altaffiltext{2}{Now at Rochester Institute of Technology, 84 Lomb Memorial Drive, Rochester NY 14623-5603 USA}

\email{clbsps@cis.rit.edu, peter,geoff,mad@mso.anu.edu.au}

\shorttitle{Radio-Excess \iras\ Galaxies}
\shortauthors{Buchanan et al.}

\begin{abstract}
This is the fourth in our series of papers discussing the nature of
radio-excess galaxies, which have radio emission associated with an active
nucleus but which do not fit into the traditional categories of either
radio-loud or radio-quiet active galaxies.  In this paper, we present optical
spectra of our sample of FIR-luminous radio-excess galaxies.  Optical emission
line ratio diagnostics are used to determine the dominant source of the
ionizing radiation. We find that radio excess is an excellent indicator of the
presence of an active nucleus. The radio-excess sample contains a much higher
fraction of AGN than samples selected on FIR luminosity alone, or using other
criteria such as warm FIR colors.  Several objects have ambiguous
classifications and are likely to be composite objects with mixed
excitation. The type of optical spectrum appears to be associated with the
radio-loudness: our results suggest that radio-loud objects may be more `pure'
AGN than radio-intermediate objects.  We find strong evidence for interaction
between the radio plasma and the surrounding gas. Broad, structured optical
emission lines are observed and a relative blueshift is measured between the
\oiiia\ and \halpha\ lines in several sources.  Jet energy fluxes are inferred
from the \oiiia\ luminosities using a shock model for the interaction between
the radio jet and the line-emitting gas. The jet energy fluxes of the
radio-excess objects are lower than in powerful radio sources, consistent with
our previous results. We conclude that the jets of radio-intermediate sources
are intrinsically weaker than those in sources with more powerful radio
emission.  A significant fraction of the sample spectra show post-starburst
stellar continuum, with A-star absorption lines. Post-starburst stellar
populations are consistent with the large fraction of merging or disturbed
host galaxies in the sample. The ages of the radio sources are significantly
less than those of A stars indicating that, if the radio sources are
associated with merging activity, there is a delay between the interaction and
the initiation of the radio activity.
\end{abstract}


\section{Introduction}
\label{spe_sec:int}

Optical emission lines are a characteristic feature of active galaxies.  The
ionizing source for the optical emission lines seen in AGN is a combination of
photoionization by ultraviolet (UV) continuum emission from the nucleus and
shock ionization produced by high velocity outflows such as radio jets. In
powerful radio galaxies, the optical emission lines reveal the interaction
between the radio plasma and the surrounding medium.  Detailed studies of
individual radio galaxies suggest that the radio and line-emitting regions are
spatially coincident and reveal kinematic evidence for shocks and acceleration
of the line-emitting material by the jet and overpressured lobes (e.g.,
\citealt{bre84a, bre84b, tad91, koe98, vil99, sol01a, sol01b, koe02}).

Studies of samples of radio galaxies have shown that the interaction between
the radio plasma and the surrounding medium is related to the radio source
size (\citealt*{bes00}, \citealt{bes02, ins02b}). These results have shown
that radio plasma/cloud interactions dominate the kinematics and ionization
of the line-emitting material while the source is young and expanding through
the interstellar medium of the host galaxy. In more evolved radio sources,
with larger extents, the line-emitting material is predominantly photoionized
by the central ultraviolet continuum source, and the kinematics are consistent
with gravitational motion in the host galaxy potential.  The radio plasma
interactions may take the form of direct jet--cloud interactions or
interactions with the over--pressured radio lobe.

Powerful compact radio sources that are embedded in the dense inner regions of
the host galaxy show evidence for strong interaction between the expanding
radio lobes and the interstellar medium.  \citet{gel94} discovered broad,
complex line profiles in compact-steep spectrum (CSS) sources. A detailed
study of three CSS sources found broad emission line widths, complex line
profiles, and velocity offsets in the emission line gas surrounding the radio
sources \citep{ode02, ode03}.  The \oiiia\ emission line in two other CSS
sources has been shown to be blueshifted by 600 -- 2000 \kms\ with respect to
the systemic velocity of the host and low ionization lines \citep{hol03,
tad01}. These studies provide strong evidence for acceleration of the
line-emitting gas by powerful radio jets.  Some evidence has also been found
for interaction between the much weaker radio sources found in Seyfert
galaxies and the surrounding medium \citep{whi92b}.

The radio continuum and \oiiia\ line luminosities are found to be correlated
in samples of radio galaxies (e.g., \citealt{bau89, raw91, tad98,
bau00}). This suggests either a direct relation between the radio jets and the
line emission, or that both are directly related to the luminosity of the
central engine.  A correlation between the radio power and \oiiia\ emission is
also seen in Seyfert galaxies \citep{bru78, whi85}.  This correlation is
separated from the radio galaxy correlation by approximately four orders of
magnitude in radio power and this has been used to argue for a dichotomy
between radio-loud and radio-quiet AGN \citep{wil95}.  The correlation between
radio power and \oiiia\ luminosity in Seyfert galaxies is most likely due to
the relation both have with the properties of central engine, although the
physical details of this correlation remain to be elucidated.

In this paper we investigate the optical spectroscopic properties of our
sample of 49 FIR-luminous radio-excess galaxies, and the hypothesis that they
resemble low-power \cg\ sources. The spectroscopic observations and
measurements are described in \S\ref{spe_sec:obs}. The line ratios and object
classifications as well as line widths and equivalent widths are presented in
\S\ref{spe_sec:res}. In \S\ref{spe_sec:dis} we discuss the classifications and
the \oiiia\ luminosities of the radio-excess objects as a function of their
radio power. We use models to infer properties of the radio jets from the
optical line emission, and discuss the presence of stellar continuum and
absorption lines in sample spectra. Our conclusions are presented in
\S\ref{spe_sec:con}.  For consistency with our previous work, we adopt the
cosmological parameters \ho\ = 50 \kms Mpc$^{-1}$ and \qo\ = 0.5.

\section{Optical spectroscopy}
\label{spe_sec:obs}

The selection of the radio-excess sample is described in detail in
\citet[hereafter \paperone]{dra03}.  Briefly, objects were found by
cross-correlating the Parkes-MIT-NRAO 5 GHz radio source catalog \citep{gri93,
gri94, wri94, gri95, wri96} with the \iras\ FSC \citep{mos92}. Objects having
more than 5 times as much radio emission as expected from the
far-infrared/radio correlation followed by star-forming galaxies and
radio-quiet active galaxies were selected to form the radio-excess sample.
This criterion corresponds to \uexpr\ $<$ 1.80.  The FIR-luminous subsample
selected for further study comprises those radio-excess objects with \nLnfir\
$> \: 10^{11}$ \lsun. This selection is described in \citet[hereafter
\papertwo]{dra04a}.  The FIR-luminous radio-excess objects were
observed with the Double-Beam Spectrograph (\dbs) on the Australian National
University (ANU) 2.3 m telescope at Siding Spring Observatory. The B300 and
R316 gratings were used in the blue and red arms, respectively, giving a
spectral resolution of $\sim$~5~\AA. The observations and reduction of the
spectra are described in detail in \paperone.  The low resolution optical
spectra are shown in Appendix \ref{spe_sec:app_spec}, Figures
\ref{spe_fig:spec}\emph{a} - \emph{m}. Some of the objects were observed again
with the B600 and R600 gratings of the \dbs, at a spectral resolution of
$\sim$2 \AA. These spectra were reduced by the same method as the lower
resolution spectra.

Spectrophotometric measurements were performed on the lower resolution
spectra using the \iraf\ data reduction package. The line fluxes and
widths of unblended lines were measured by Gaussian fitting using the
\splot\ task. At the resolution of the observations, several emission
lines were blended, including \halpha\ with \niiab\ and \siia\ with
\siib. Parameters for these lines were measured by simultaneously
fitting multiple Gaussians using the task \ngaussfit. This task allows
the widths and fluxes of lines of the same species to be fixed
relative to each other. Where \hbeta\ was present in both absorption
and emission, \ngaussfit\ was used to fit these simultaneously.  For
eight objects the \halpha\ and \hbeta\ emission was fitted with a
broad as well as a narrow component.  The line flux in these cases is
listed as the sum of the two components but the line ratios were
calculated using the narrow component only.

The line fluxes were corrected for reddening following \citet{vei87}.  The
\halpha\ and \hbeta\ lines were used to determine the Balmer decrement, and
the Whitford reddening curve was used to determine the reddening
\citep{mil72}.  An intrinsic flux ratio of \halpha/\hbeta\ $ = 3.1$ was
assumed for all objects. This value is slightly higher than the Case B Balmer
recombination ratio of 2.85 expected for starburst galaxies and is more
appropriate for active galaxies \citep{vei87}, however we note that the object
classifications (\S\ref{spe_sec:res}) are unchanged if an intrinsic
\halpha/\hbeta\ $ = 2.85$ is used for the reddening correction.  The
(uncorrected) \halpha\ and \hbeta\ fluxes, along with the derived extinction,
are listed in Table \ref{spe_tab:lines_a}.  The fluxes of other emission
lines, corrected for reddening, are listed in Table
\ref{spe_tab:lines_b}. Where \hbeta\ is an upper limit, the reddening estimate
is a lower limit.  Two spectra had no \halpha\ emission line; no reddening was
calculated for these spectra.

The line widths were taken to be the full width at half-maximum (FWHM)
of the fitted Gaussian. The instrumental broadening was measured by
fitting Gaussian profiles to emission lines in sky spectra extracted
from the same image as the object spectrum. These were typically
$\sim$~5.0~\AA. The instrumental FWHM was subtracted in quadrature
from each measured line width to estimate the intrinsic line
width. These are listed in Table \ref{spe_tab:fwhm}. Equivalent widths
of the emission lines were derived from the fitted Gaussian profiles
using the \splot\ task. The equivalent width of the \oiiia\ emission
line for each spectrum is listed in Table \ref{spe_tab:EQ}.

\section{Results}
\label{spe_sec:res}

\begin{figure}[H]
\epsscale{0.8}
\plotone{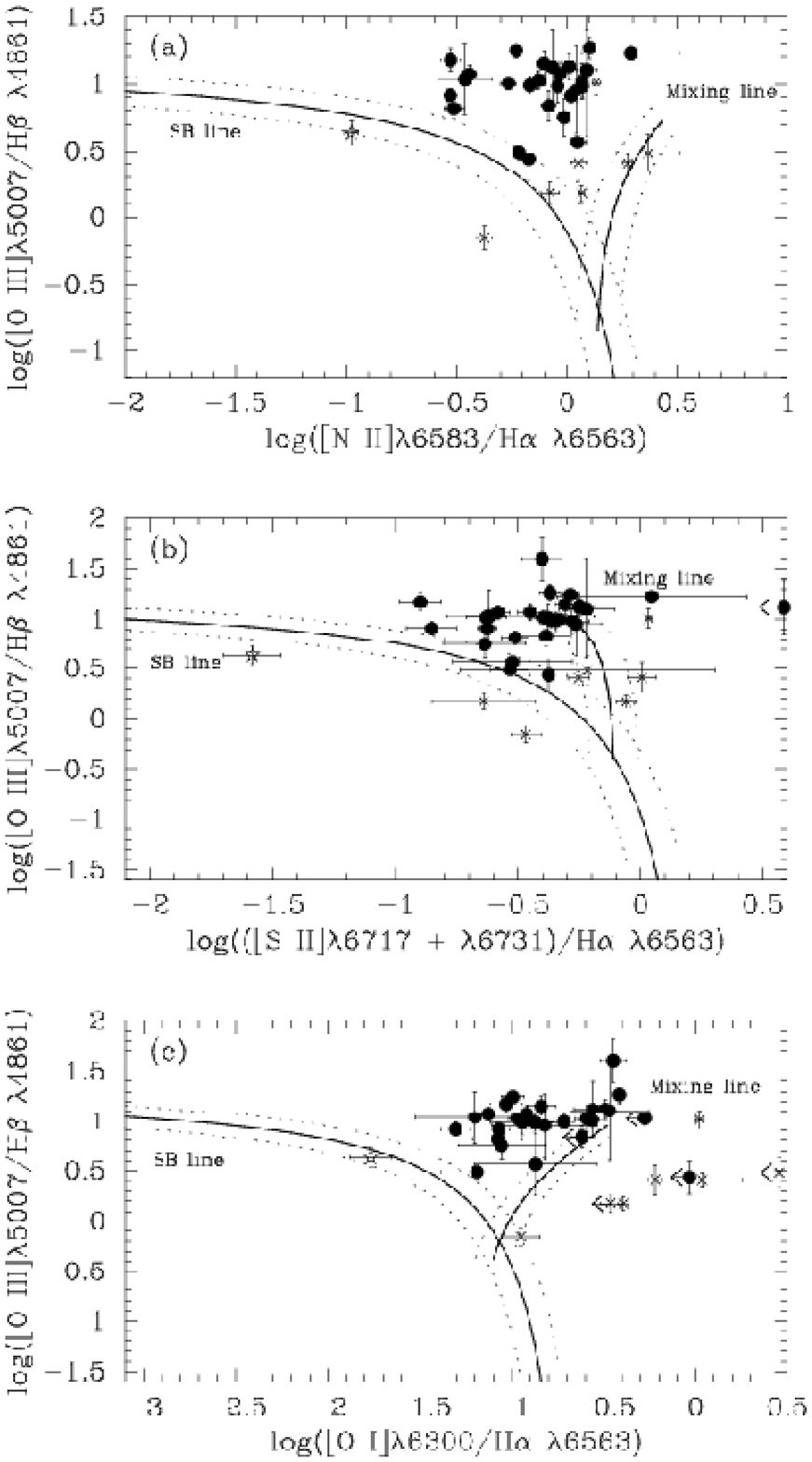}
\caption{\small Diagnostic diagrams. (a) log(\protect\oiiia/\protect\hbeta) vs
log(\protect\niia/\protect\halpha). (b) log(\protect\oiiia/\protect\hbeta) vs
log((\protect\siia + \protect\siib)/\protect\halpha). (c)
log(\protect\oiiia/\protect\hbeta) vs log (\protect\oi/\protect\halpha).
Seyfert galaxies (\emph{filled circles}), LINERs (\emph{open squares}), and
starburst galaxies (\emph{open stars}) are classified according to their
location relative to the maximum starburst line (\emph{solid line}) and
extreme mixing line (\emph{solid line}) \protect\citep{kew01}. Objects are
classified as ambiguous (\emph{x's}) if they fall in different regions in
different diagrams. Line fluxes have been corrected for
reddening.\label{spe_fig:diag}} \epsscale{1.0}
\end{figure}

Reddening-corrected narrow-line ratios were derived and plotted on
three \citet{vei87} diagnostic diagrams to determine the nature of the
line-emitting region (see Figure \ref{spe_fig:diag}).  The theoretical
scheme of \citet{kew01} was used to classify the galaxies into
starbursts (SB), Seyferts (Sy), and low-ionization narrow
emission-line region (LINER) galaxies. This scheme has been shown to
give similar results to the semi-empirical scheme of \citet{vei87},
but with fewer ambiguous classifications \citep{kew01,cor03}.  Objects
below and to the left of the maximum starburst (\emph{solid}) line on
each diagram are classified as SB. Objects above this line and to the
left of an extreme mixing line (\emph{solid line}) are classified as
Sy. The mixing line indicates the locus of objects with a maximum
starburst combined with 0 -- 100\% contribution by an AGN component.
Objects above the maximum starburst line and to the right of the
extreme mixing line are classed as LINERs. Objects were classified by
their location in the three diagnostic diagrams. In most cases the
three classifications agreed and a final type was assigned to the
object. Objects for which the three classifications did not agree are
described as ambiguous.  The classifications and assigned types are
listed in Table \ref{spe_tab:class}.

The diagnostic diagrams (Figure \ref{spe_fig:diag}) reveal that the
majority of the objects have Seyfert spectra.  We find 27 galaxies
with Seyfert spectra (75\%) and one galaxy with a starburst spectrum
(3\%). No objects were unambiguously classified as LINERs. The
remaining eight objects (22\%) had ambiguous classifications. Of
these, five were Sy/LINER, one was SB/LINER and one was Sy/SB/LINER.
The Seyfert galaxies were further divided into broad- and narrow-line
objects (Seyfert types 1 and 2) based on the \halpha\ line widths
(Table \ref{spe_tab:fwhm}). Seyferts with FWHM$_{\halpha} > 2000$
\kms\ were classified as Sy 1 and objects with narrower \halpha\
emission lines as Sy 2.

We extracted fixed-size apertures for all sample objects.
\citet{vei02a} noted that aperture effects can cause line ratios to
move towards the starburst region of the diagram, when the same sized
aperture is used for objects of different redshifts. This occurs
because the corresponding linear aperture is larger for higher
redshift objects and starburst regions often surround the nucleus. The
potential contamination from any circumnuclear starburst increases
with redshift. The extracted apertures (12 pixels $\sim$ 10\arcsec)
correspond to scales of $\sim$~10 -- 70~kpc at the redshifts of our
objects. However we see no evidence that objects with higher redshifts
lie closer to the starburst region than lower redshift objects.

\section{Discussion}
\label{spe_sec:dis}

\subsection{Classifications}
\label{spe_subsec:dis_clas}

Classifying objects based on the diagnostic diagrams, we find the
majority of objects in the sample (75\%) are AGN. Using the
theoretical scheme of \citet{kew01}, we find that 56\% of the objects
are Seyfert type 2 and 19\% are Seyfert type 1 (FWHM$_{\halpha} ~ >$
2000~\kms). Only one object is classified as a starburst and no object
is classified as a LINER. Classifications for eight objects (22\%)
were ambiguous. These objects are discussed further below.

We find a higher fraction of AGN than in samples of objects selected
on FIR luminosity alone, which are dominated by radio-quiet objects at
\iras\ flux densities. This confirms that the radio-excess criterion
selects AGN (see \paperone).  Studies with other selection criteria,
aimed at selecting AGN, have also found a higher fraction of AGN,
though none of these contain as high a fraction as found in our
sample. For example, \citet{gri92} selected IR galaxies with warm FIR
colors and found 22\% Seyfert 1, 39\% Seyfert 2 galaxies and 37\%
starburst galaxies.  Samples selected only on FIR luminosity find
lower fractions of AGN and higher fractions of starburst and LINER
galaxies.  In their sample of luminous infrared galaxies,
\citet{vei95} found 12\% of their objects were Seyfert 2 and only 2\%
were Seyfert 1, while 59\% of the sample were SB and 27\% were LINER
galaxies.  \citet{kew01} found a similarly low fraction of AGN and
high fraction of starbursts in their sample of largely radio-quiet
IR-luminous galaxies. They find only 0.4\% of LINERs in their sample.
Some of this difference may also be attributable to different
classification methods, especially for the LINERs.  \citet{vei95} use
the criterion \lratioOIIIHb\ $\leq$ 3 to classify LINERs, while
\citet{kew01} use the criteria we adopted.  \citet{kew01} note that
6\% of galaxies in their sample are classified as LINERs using the
criteria of \citet{vei95}. We find that five objects with ambiguous
classifications and one object classified as a Seyfert, in total 12\%
of our sample, have \lratioOIIIHb\ $\leq$ 3.

\begin{figure}[H]
\epsscale{0.8}
\plotone{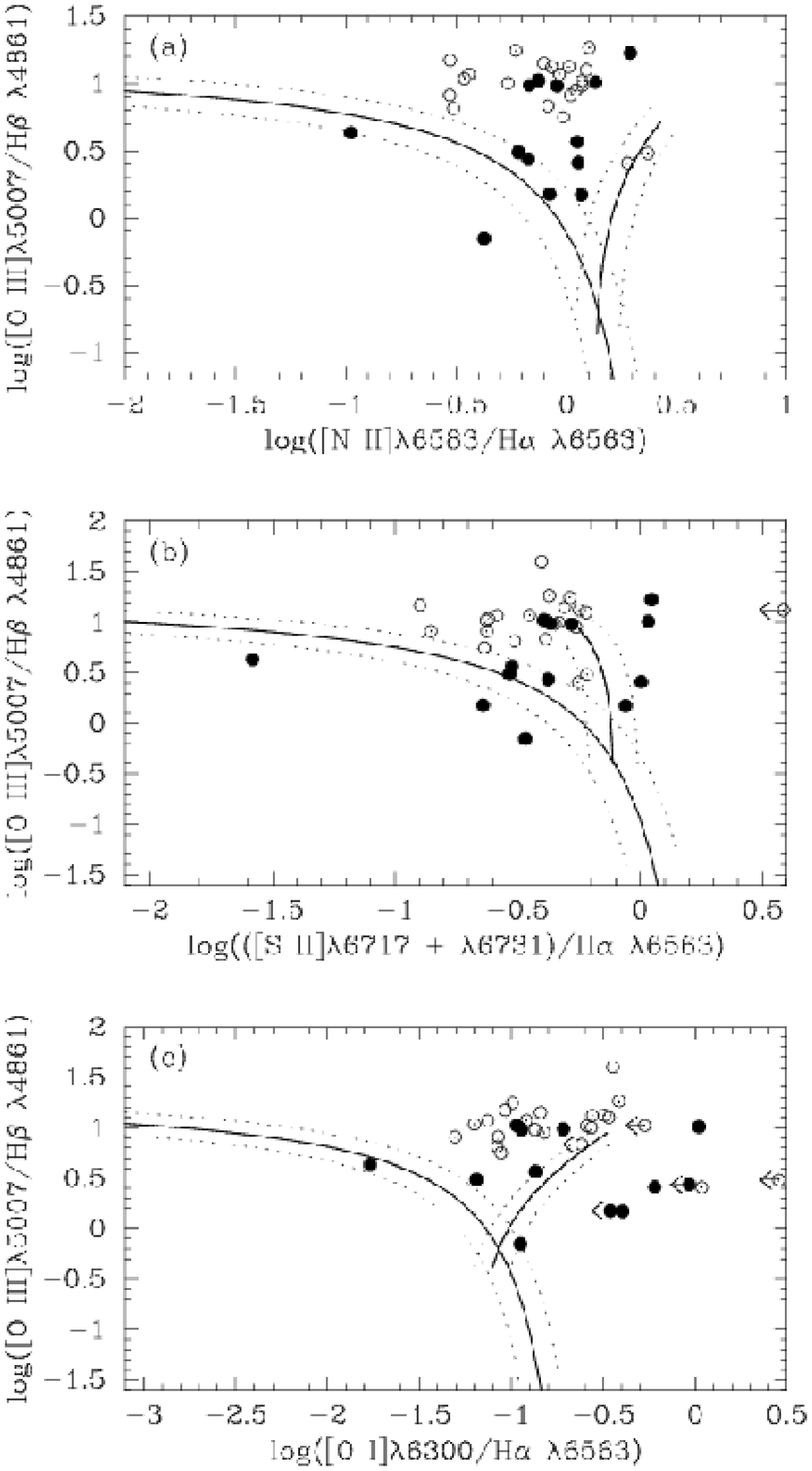}
\caption{\small Diagnostic diagrams. The symbols indicate the level of radio
excess, radio-loud (\emph{open circles}) or radio-intermediate (\emph{filled
circles}).\label{spe_fig:diag_u}} \epsscale{1.0}
\end{figure}

Figure \ref{spe_fig:diag_u} shows the diagnostic diagrams with symbols
indicating the level of radio-excess of the object. Following
\paperone, we have divided the sample into radio-intermediate (0.0$\,
< \, u \, < \,$1.80) and radio-loud ($u \, < \,$0.0) objects, where
the radio-excess parameter $u \, = \, \log \, [$\Snfir $/$ \Snrad$]$.
There is a clear tendency for the radio-loud objects (\emph{open
circles}) to be higher up the mixing line derived by \citet{kew01} and
so exhibit a higher percentage of pure AGN excitation than the
radio-intermediate objects.  The Kaplan-Meier estimator shows the
means of the \lratioOIIIHb\ for the radio-intermediate and radio-loud
objects to be 0.78$\pm$0.07 and 1.05$\pm$0.07, respectively. However,
two-sample Gehan and logrank tests using survival analysis methods to
account for lower limits indicate differences in \lratioOIIIHb\
between the radio-intermediate and radio-loud objects with
significance levels of only $\sim$10\%. A larger sample is required to
investigate properly the relation between the radio loudness and the
properties of the optical spectrum. The radio-intermediate objects are
less powerful AGN in both the radio and the optical, compared with the
radio-loud objects, so it may be that the radio-intermediate objects
have a composite excitation containing a higher fraction of starburst
contribution.

One object is classified as starburst on the diagnostic diagrams.  This
object, F21139-6613, falls quite close to the maximum starburst line in Figure
\ref{spe_fig:diag}, within the uncertainties, in two of the three diagrams.
It has a modest radio excess ($u\:$=$\:$1.2) and high FIR luminosity ($\log$
\nLnfir\ = 11.7 \lsun). It has the highest extinction of any object in the
sample, with E(B-V) = 3.1.  It is therefore likely to be a composite
starburst-AGN in which the AGN is highly reddened and consequently obscured at
optical wavelengths.

Several of the ambiguous classifications can be redefined as
borderline rather than ambiguous, as the objects lie very close to the
borders between regions in one or more diagrams. The eight objects
with ambiguous classifications tend to have large uncertainties in
their line ratios. These objects may also be composite starburst/AGN
in which nuclear activity is either weak or highly obscured.

\begin{figure}[H]
\epsscale{0.8}
\plotone{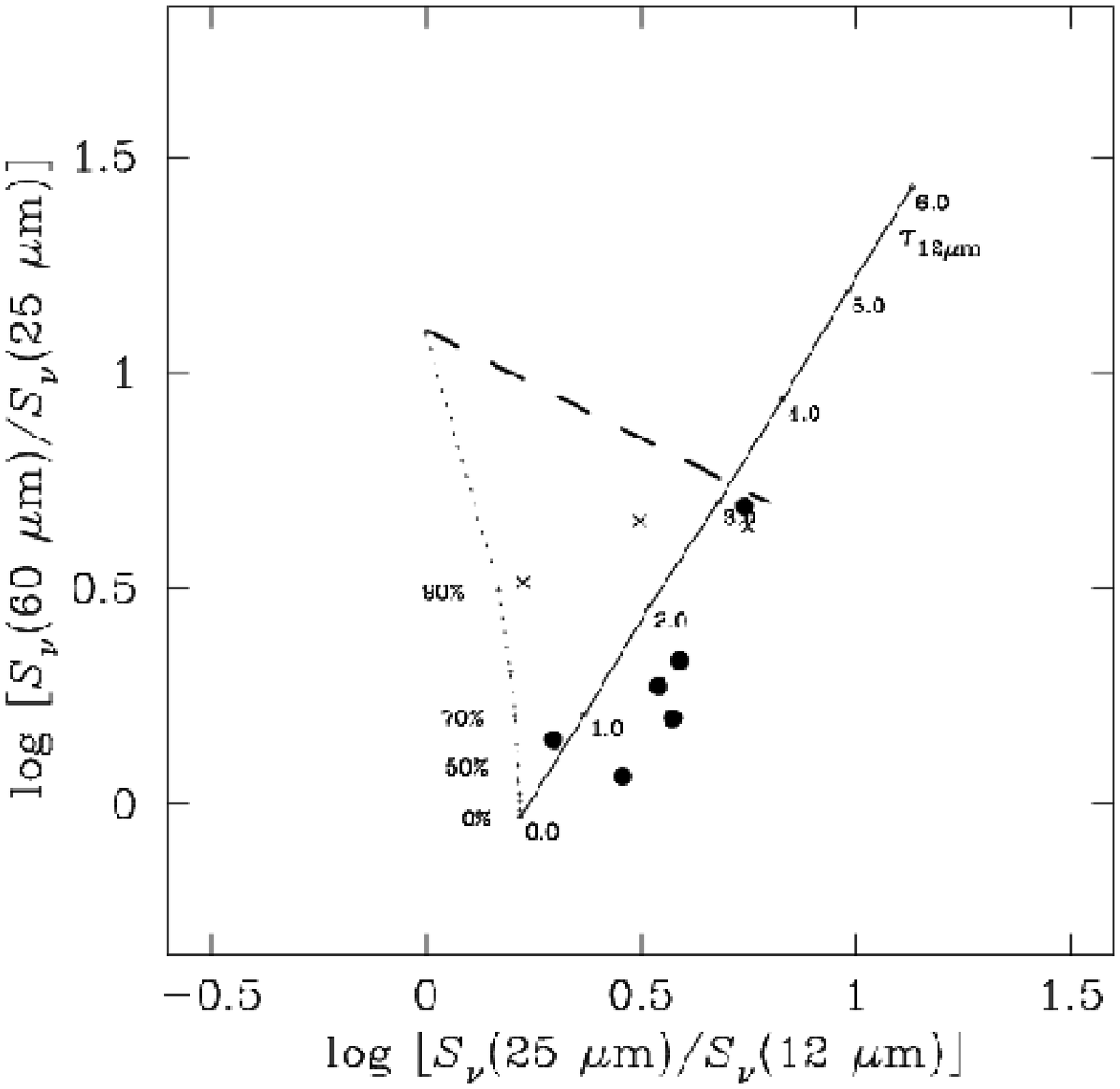}
\caption{\small FIR color-color diagram for objects with measured \iras\ 25
  \micron\ and 12 \micron\ fluxes. The symbols indicate the
  classification assigned to the objects based on the diagnostic
  diagrams, Seyfert (\emph{filled circles}) or ambiguous
  (\emph{x's}). The loci marked on the diagram follow
  \protect\citet{dop98}. The solid line is a reddening line produced
  by increasing the optical depth, $\tau_{\, 12 \micron}$, through
  which a pure Seyfert nucleus is observed. The dotted line is a
  mixing line indicating the location of objects comprising a mixture
  of pure AGN (0\% starburst) and increasing fractions of
  starburst. The dashed line indicates the location of starburst
  galaxies.\label{spe_fig:FIRcols}}
\epsscale{1.0}
\end{figure}

Figure \ref{spe_fig:FIRcols} shows the FIR color-color diagram for the
objects with measured \iras\ 25 \micron\ and 12 \micron\ fluxes
(\paperone). While the majority of objects with unambiguous
classifications do not have measured \iras\ fluxes in these bands, the
few objects that can be plotted are mostly Seyferts that fall along
the reddening line \citep{dop98}. Thus the optical diagnostics and FIR
colors of these objects agree, suggesting that these are highly
reddened Seyfert galaxies.  The location of the three ambiguous
objects on this diagram is consistent with these being composite
and/or highly obscured objects.  

\subsection{Jet/ISM Interaction}
\label{spe_subsec:dis_int}

\begin{figure}[H]
\epsscale{0.8}
\plotone{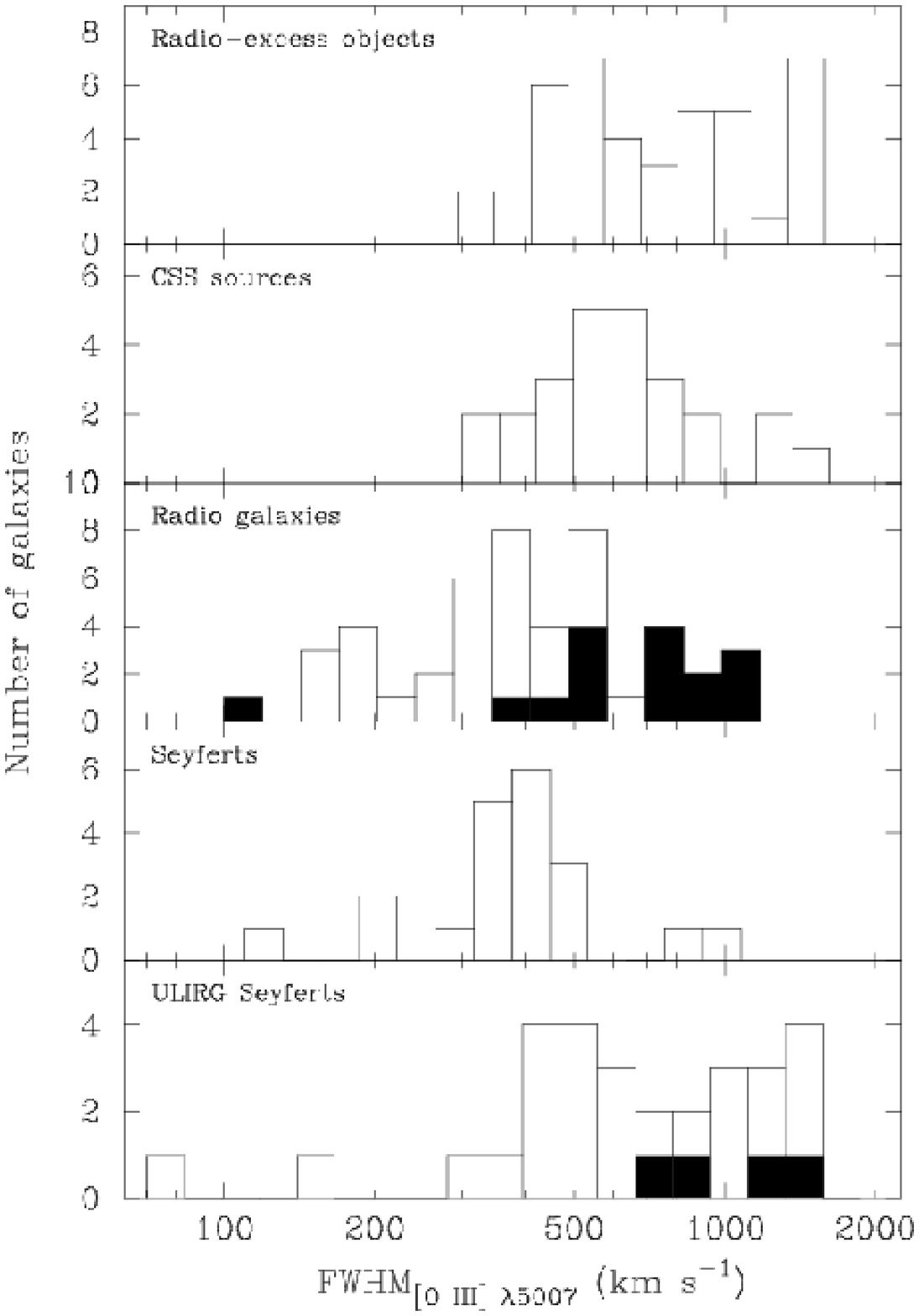}
\caption{\small The distribution of \protect\oiiia\ emission line widths for
the FIR-luminous radio-excess sample and comparison objects. CSS
sources are from \protect\citet{gel94}, radio-galaxies from
\protect\citet{bau00}, Seyfert galaxies with linear radio sources are
from \protect\citet{whi92a}, and ULIRG Seyferts are from
\citet{vei99}. Filled regions are explained in the
text.\label{spe_fig:histOIIIfwhm}}
\epsscale{1.0}
\end{figure}

Most of the FIR-luminous radio-excess objects show broad \oiiia\ line
widths. The distribution of measured line widths for our objects, corrected
for instrumental broadening, are shown in Figure \ref{spe_fig:histOIIIfwhm}
(\emph{top panel}). Data from the literature for other objects are shown for
comparison.  The line widths of the FIR-luminous radio-excess galaxies cover a
broad range of velocities from as narrow as $\sim$ 250 \kms\ to almost 2000
\kms.  The line widths are broader than typical galaxian rotation, suggesting
that there is strong interaction between the radio source and the
line-emitting gas in the radio-excess objects.  The line widths are similar to
CSS sources and broader than observed in extended radio galaxies and Seyferts
with linear radio sources.

Broad line widths have been observed in powerful compact radio sources and
interpreted to indicate strong interaction between the radio jet and the host
ISM (e.g., \citealt{gel94, ode02, ode03}). \citet{whi92b} has shown that the
line widths of radio-quiet Seyfert galaxies correlate with the rotational
velocity of the galaxy, and that Seyfert galaxies with linear radio sources
have broader line widths. The kinematics of the narrow-line region in
radio-quiet galaxies therefore seems to be dominated by the galaxy potential.
The cited work indicates that in those objects with linear radio sources, the
interaction of the radio plasma with the line-emitting material accelerates
the gas and produces broader line widths.  In the case of powerful radio
galaxies, the line widths are not as broad as in CSS sources because the radio
jets do not interact as strongly with the gas in the narrow-line region. The
powerful radio galaxies show similar line widths to the Seyfert
galaxies. However, all except one of the high-redshift radio galaxies ($z >
0.6$; \emph{filled histogram, middle panel}) fall at the upper end of the
distribution, with the broadest line widths. \citet{bau00} note this
difference between the low- and high-redshift objects in their sample. If
broad line widths in these objects are attributable to interaction between the
radio source and the surrounding gas, the increase in line width with redshift
can be interpreted as an increase in strength of the interaction with radio
power. Alternatively, \citet{bau00} argue that if the broad line widths are
associated with larger gravitational potentials, the increase in line width
with redshift indicates that the environment of radio galaxies changes with
redshift.  Thus part of the correlation may simply be an environmental effect.
Recent studies suggest that both effects are operating; radio sources interact
more strongly with the environment when they are small and expanding into the
dense host galaxy medium, and more extreme kinematics are observed at higher
redshift as a result of different environments \citep{bes00, bes02, ins02a,
ins02b}.

ULIRG Seyferts show a large range of line widths, some as broad as those in
our sample. Little is known about the radio properties of these objects, but
at least four are known to be radio-excess objects, including two objects that
are also present in our sample. These four objects all have broad line widths
(Figure \ref{spe_fig:histOIIIfwhm}; \emph{filled histogram, bottom
panel}). The ULIRG Seyferts would seem to comprise both radio-quiet objects in
which the \oiiia\ line width reflects the galaxian rotation, and radio-excess
objects, in which there is additional acceleration of the line-emitting
material by the radio plasma.

\begin{figure}[H]
\epsscale{0.8}
\plotone{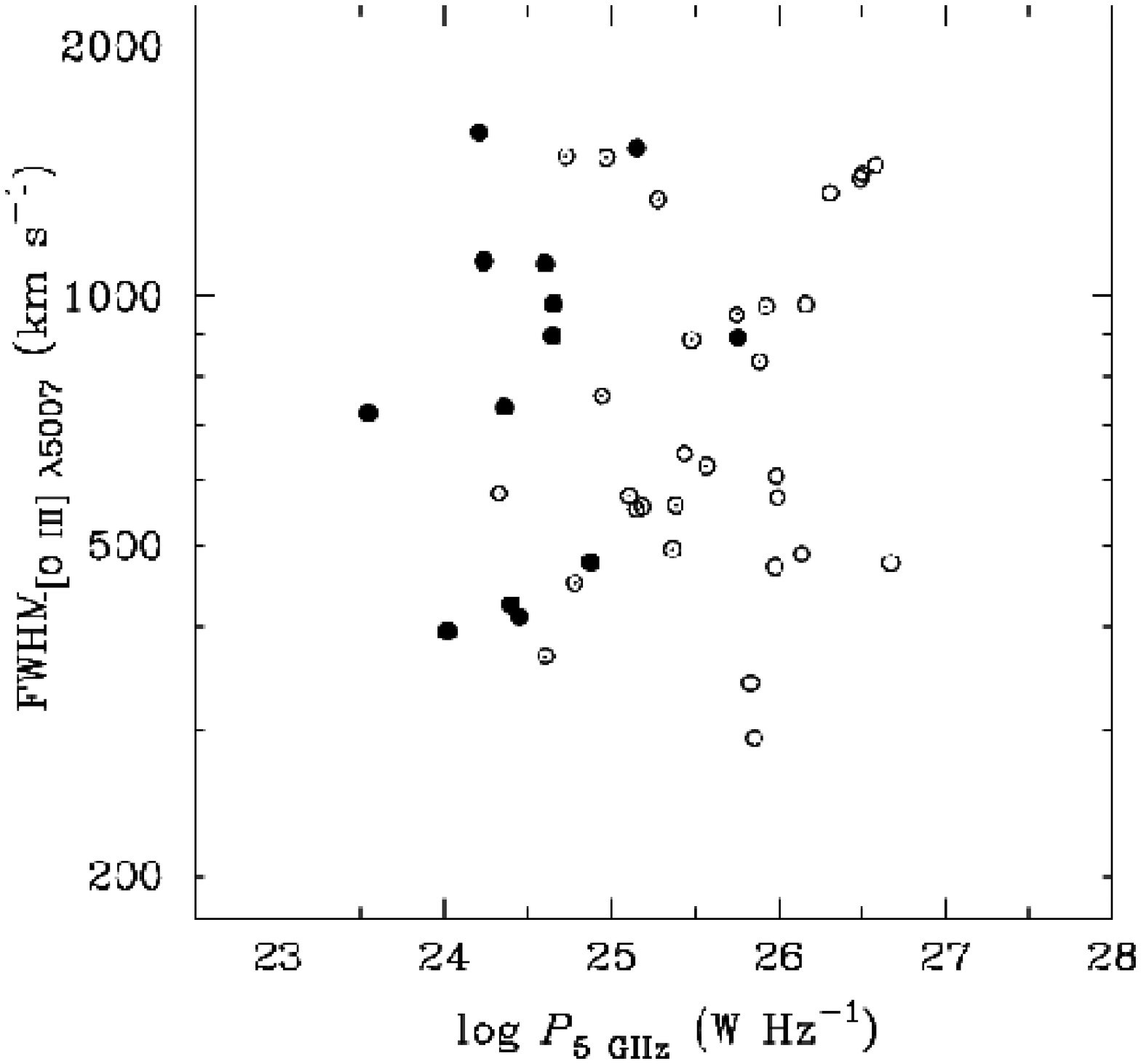}
\caption{\small The \protect\oiiia\ emission line width relative to the radio
power. The symbols indicate the level of radio excess: $u ~ <$ 0.0
(radio-loud; \emph{open circles}), 0.0 $< ~ u ~ <$ 1.0 (\emph{dotted
circles}), and 1.0 $< ~ u ~ <$ 1.8 (\emph{filled circles}). We note
that the apparent correlation between the radio power and radio excess
is a selection effect in our sample.\label{spe_fig:OIIIfwhm_rad}}
\epsscale{1.0}
\end{figure}

We do not detect any correlation of the \oiiia\ line width with the radio
power (Figure~\ref{spe_fig:OIIIfwhm_rad}). This is consistent with the similar
line widths seen between the radio-excess galaxies and the more powerful \cg\
sources and suggests that the radio power is not the dominant factor
determining the amount of line broadening. Other factors, such as the amount
of gas in the narrow-line region, may be more important.

\begin{figure}[H]
\epsscale{0.7}
\plotone{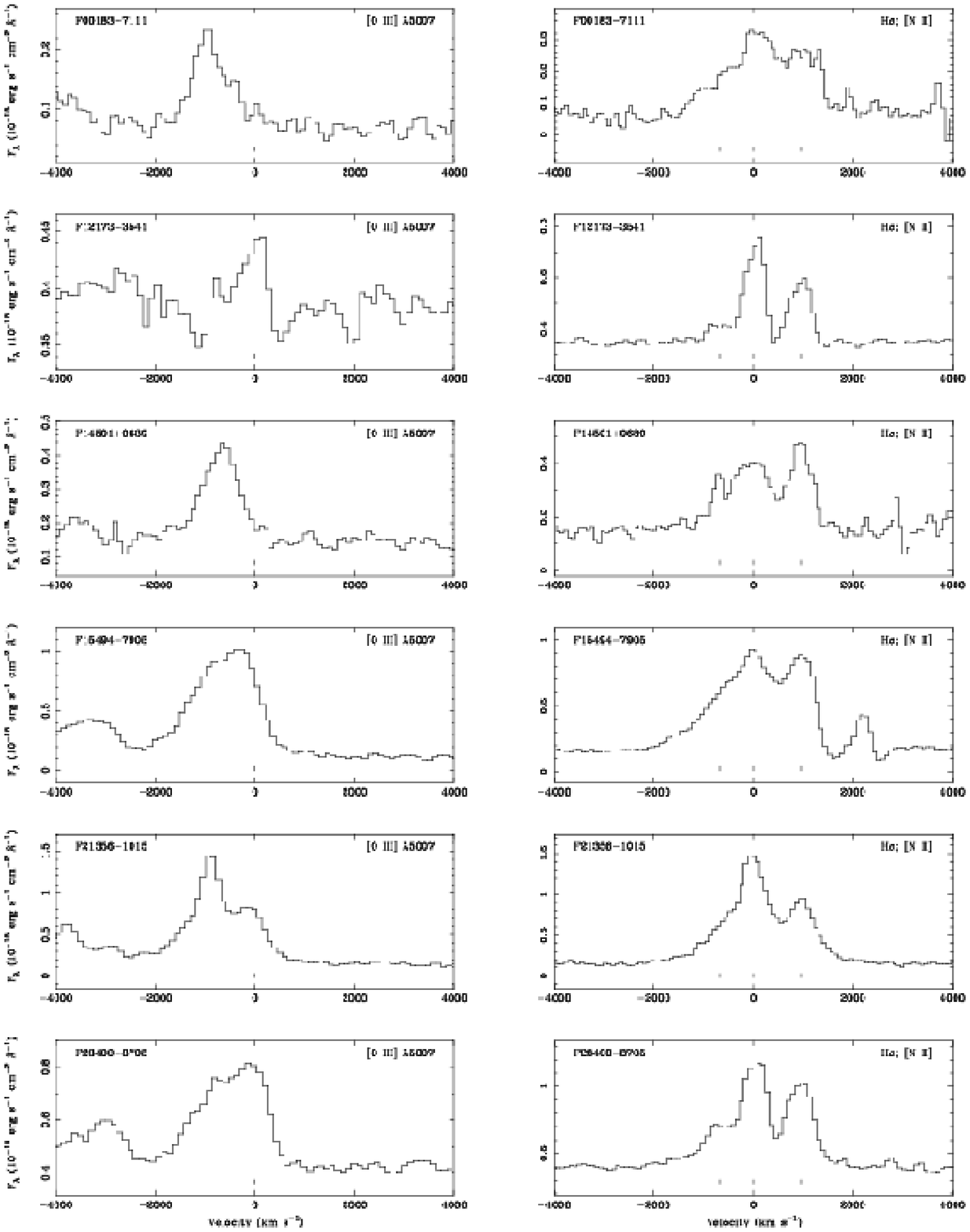}
\caption{\small Plots of the spectra of six objects showing a blueshift
  between the \protect\oiiia\ and \protect\halpha\ emission
  lines. Each line shows one object, with the left and right panels
  showing the \protect\oiiia\ and \protect\halpha\ lines,
  respectively. The vertical scale is in units of 10$^{-15}$
  \protect\ergscA, and the horizontal scale in units of
  \protect\kms. \label{spe_fig:profblue}}
\epsscale{1.0}
\end{figure}

\begin{figure}[H]
\epsscale{0.8}
\plotone{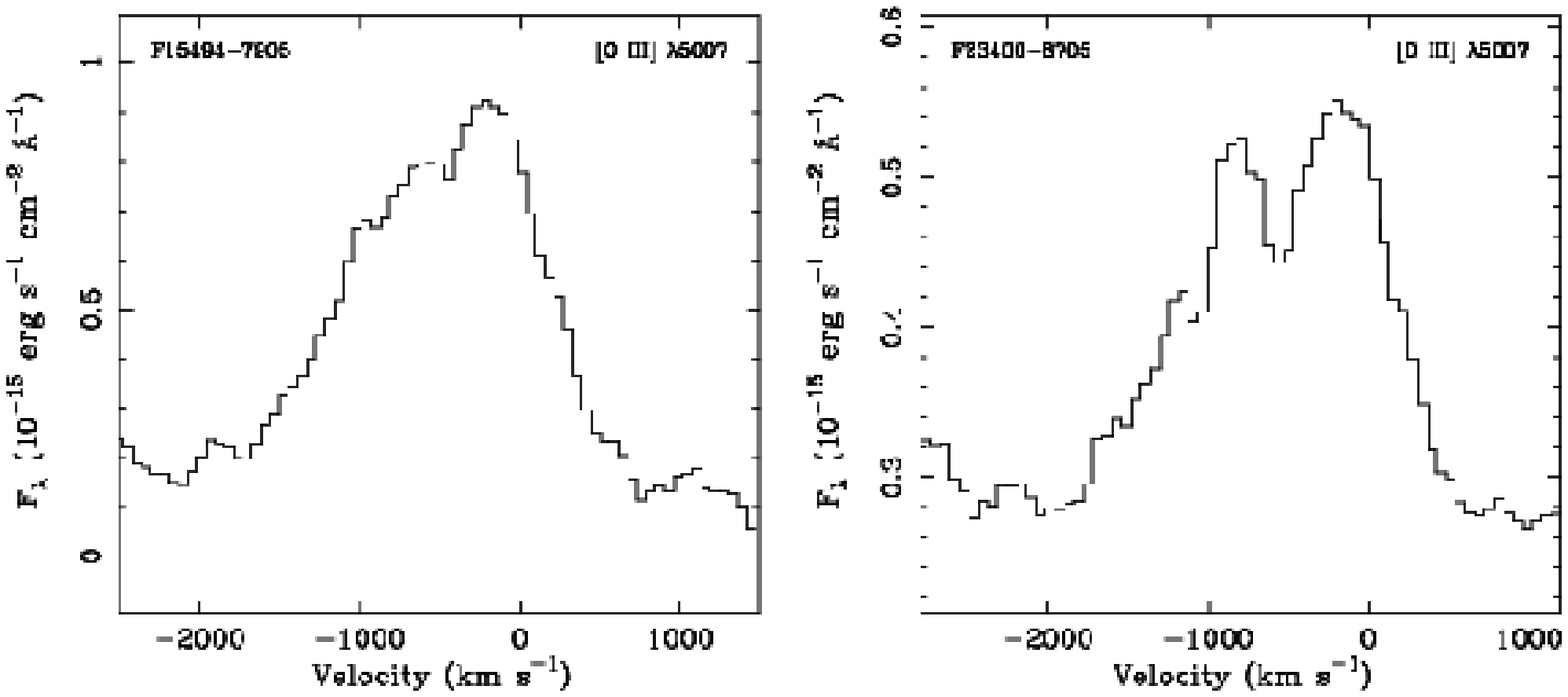}
\caption{\small Plots of the \protect\oiiia\ line profiles of two objects
  observed at higher resolution, showing the complex structure. The
  vertical scale is in units of 10$^{-15}$ \protect\ergscA, and the
  horizontal scale in units of \protect\kms. \label{spe_fig:profs}}
\epsscale{1.0}
\end{figure}

We find further evidence for acceleration of the line-emitting material in
some objects.  In six galaxies we detect a blueshift between the \oiiia\ and
\halpha\ emission lines.  These are shown in Figure \ref{spe_fig:profblue}.
Blueshifts between forbidden and permitted lines have previously been observed
in two objects in our sample that have compact powerful radio sources,
including F15494-7905, shown in Figure \ref{spe_fig:profblue} \citep{hol03,
tad01}. They have been interpreted as evidence for outflows, where the
permitted lines are at the systemic velocity of the galaxy and the forbidden
lines are produced in gas that is being accelerated and shock-excited by the
radio jet or expanding lobe.  The objects with blueshifts all have broad
\oiiia\ line widths, but they have radio powers and radio excesses covering
the full range in the sample. Thus, such outflows are not associated
exclusively with objects having high radio powers.  Complex line profiles are
also seen in some objects. Figure \ref{spe_fig:profs} shows the \oiiia\
profiles of two objects observed at higher resolution.  Complex line profiles
have been observed in powerful radio sources and Seyfert galaxies with linear
radio sources and indicate that the radio jet that is accelerating the
surrounding material produces complex kinematics in the line-emitting regions
\citep{gel94,whi92b}.

\subsection{Jet Energy Fluxes}
\label{spe_subsec:dis_jet}

\begin{figure}[H]
\epsscale{0.8}
\plotone{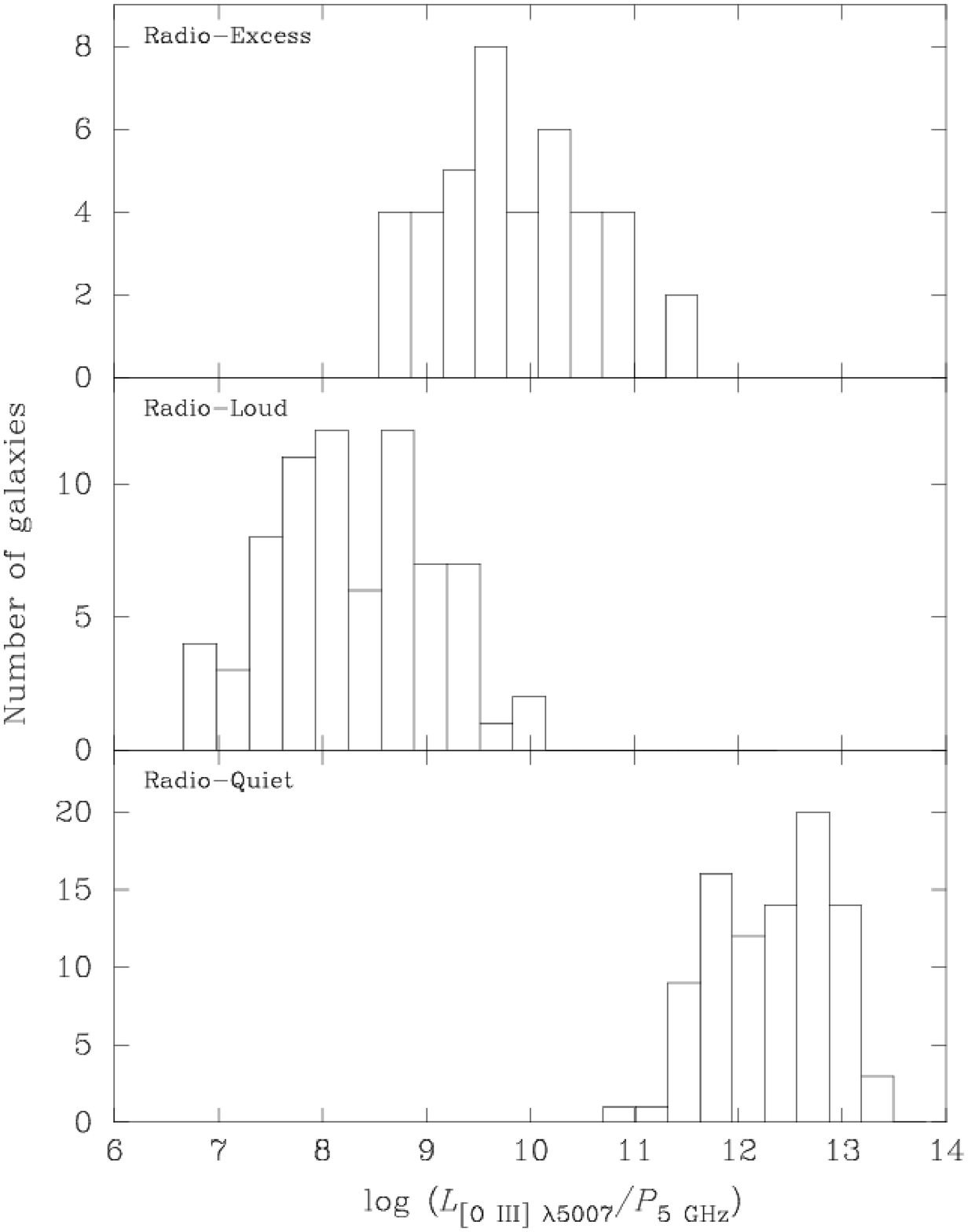}
\caption{\small The distribution of the ratio of \protect\oiiia\ luminosity to radio
power for the radio-excess sample (\emph{top panel}), for powerful radio
sources (\emph{middle panel}), and for radio-quiet Seyfert galaxies
(\emph{bottom panel}). The radio-loud objects include radio galaxies from
\protect\citet{con95, tad93, wal85, kuh81} and CSS sources from
\protect\citet{gel94}.  The data for the radio-quiet galaxies are from
\protect\citet{con95, whi92a, ede87a}.\label{spe_fig:histOIIIrad}}
\epsscale{1.0}
\end{figure}

The radio-excess objects were selected in such a way that they include many
objects that lie between radio-loud and radio-quiet objects in their radio/FIR
flux density ratios. We find that many of these are also intermediate in their
ratio of radio to \oiiia\ line luminosity.  Figure \ref{spe_fig:histOIIIrad}
shows the distributions of this ratio for the FIR-luminous radio-excess sample
and radio-loud together with radio-quiet objects from the literature for
comparison.  The radio-excess objects fall between powerful radio sources and
Seyfert galaxies in their ratio of \oiiia\ luminosity to radio power, filling
the gap between the distributions of the radio-loud and radio-quiet objects.

\begin{figure}[H]
\epsscale{0.8}
\plotone{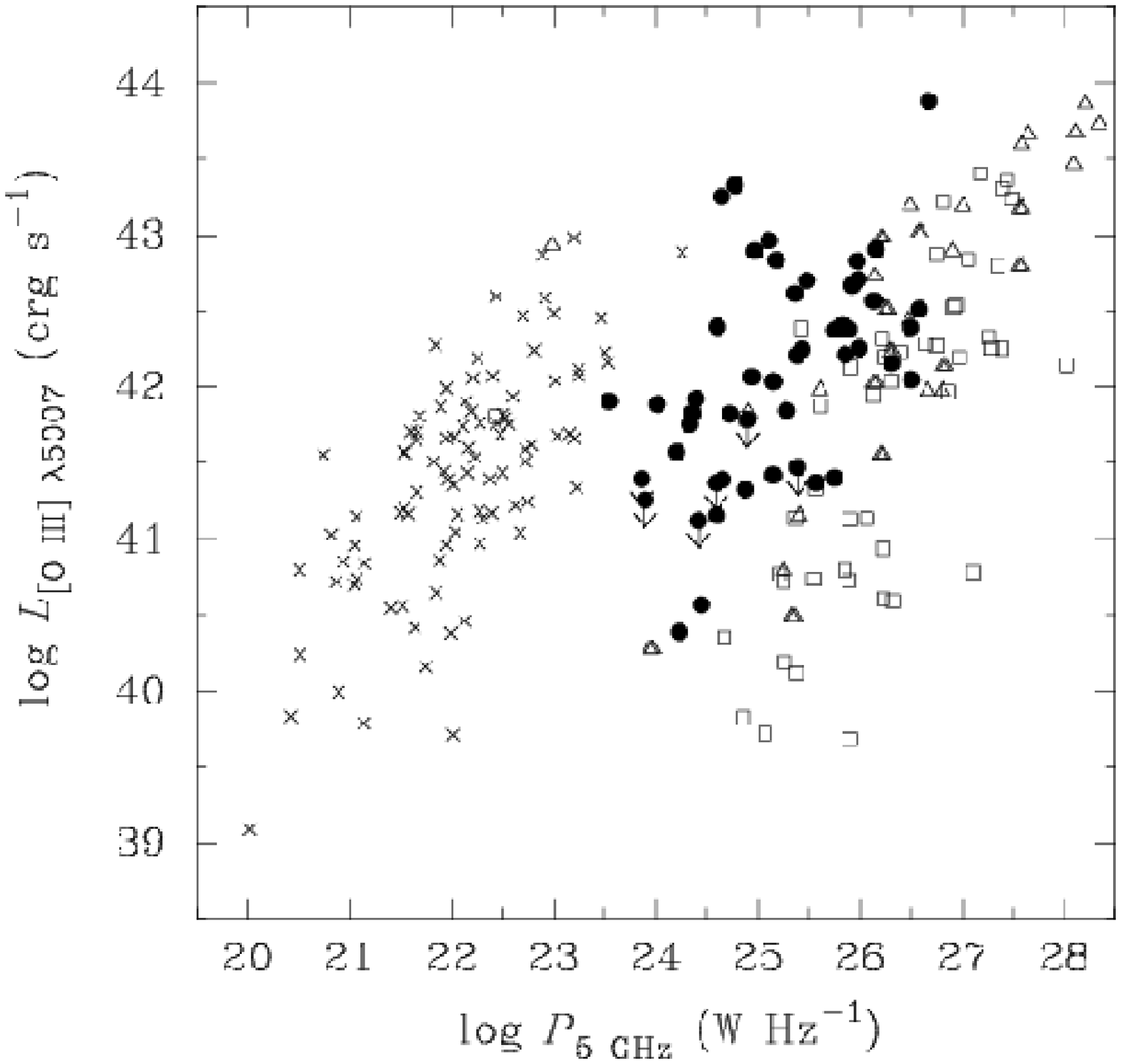}
\caption{\small The \protect\oiiia\ emission line luminosity relative to the
radio power for the FIR-luminous radio-excess sample objects
(\emph{filled circles}). The comparison objects comprise Seyfert
galaxies (\emph{x's}), radio galaxies (\emph{open squares}), and
CSS sources (\emph{open triangles}). References for the comparison
objects are given in Figure
\protect\ref{spe_fig:histOIIIrad}.\label{spe_fig:OIII_P5}}
\epsscale{1.0}
\end{figure}

We find that the radio-excess galaxies have \oiiia\ emission lines as luminous
as those in powerful radio sources, but that they lie between radio-loud
objects and radio-quiet objects on the $L_{\rm{[O III]}}$ - \pfive\ diagram.
Figure \ref{spe_fig:OIII_P5} shows the \oiiia\ emission line luminosity
relative to the radio power for the radio-excess sample and the comparison
radio-quiet and radio-loud objects.  Independent correlations have been
observed between the radio power and \oiiia\ luminosity in radio-quiet
\citep{bru78, whi85} and radio-loud sources \citep{bau89, raw91, tad98,
bau00}.  Two hypotheses to explain these correlations have been investigated:
photoionization by a central UV continuum source, or shock ionization by the
radio jets. Evidence for both has been found. \citet{bau00} found that the
emission-line luminosities in their sample of 52 3CR radio galaxies were
proportional to the nuclear UV continuum flux densities, suggesting that
photoionization is the dominant ionization mechanism.  However, their results
did not exclude ionization by shocks.  Several detailed studies of individual
objects have found morphological similarities between radio and line-emitting
regions and kinematic evidence for shocks (e.g., \citealt{bre84a, bre84b,
tad91, koe98, vil99, sol01a, sol01b, koe02}).  The alignment of line-emitting
regions with the axis of the radio jet in high-redshift radio galaxies has
been interpreted as evidence for an anisotropic UV source and for
shock-induced continuum by the radio jet \citep{bau00, bes99}. Given that
there is evidence for both mechanisms, it is likely that the observed emission
results from a combination of photoionization and shock ionization.  A number
of studies have shown that shock ionization is more important in smaller radio
sources, confined within the optical extent of the host galaxy, while the
emission-line regions in larger sources are predominantly photoionized
\citep{bes99, bes02, ins02a, ins02b}.  The broad line widths in CSS sources
support the hypothesis that the radio jet produces radiative shocks in a dense
interstellar medium \citep{gel94}.  Thus in the compact radio-excess objects,
we expect shocks to play a large role in producing the optical line emission.
However, in radio-quiet Seyfert galaxies that have no radio jets the optical
line emission is likely to be predominantly photoionized by ultraviolet
photons from the active nucleus.  The \oiiia\ emission in the radio-excess
objects may therefore include a significant photoionized component.  In some
objects, which have strong stellar continuum, there may also be a contribution
associated with star formation.  The lack of any trend of \oiiia\ line width
with radio power (Figure \ref{spe_fig:OIIIfwhm_rad}) suggests that a
combination of ionization mechanisms is operating.  The line width is bounded
at the lower limit by the gravitational motions of the galaxy, and at the
upper limit by the condition that the shocks are radiative. The distribution
of points in Figure \ref{spe_fig:OIIIfwhm_rad} indicates that most objects are
of mixed excitation.

The radio and optical line emission of compact radio sources have been
modeled by \citet{bic97}.  The model describes the evolution of a
jet-fed radio lobe as it expands into the dense interstellar medium.
In the \citeauthor{bic98} model the total shock luminosity is equated
to the rate of work done on the surrounding medium by the expansion of
the lobe. This leads to the following relation between the \oiiia\
luminosity, $L_{\rm{\oiiia}}$, and the jet energy flux, $F_{E}$.
\begin{equation}
L_{\rm{\oiiia}} \sim 1.0 \times 10^{-2} \left(
\frac{6}{8-\delta}\right) F_E ~ ~ \rm{erg~
s^{-1}},\label{spe_eqn:LOIII}
\end{equation}
where the density of the surrounding medium varies as $R^{-\delta}$,
$R$ being the distance from the nucleus. We assume $\delta =
1.5$. This relation underestimates the jet energy flux if the medium
is clumpy.  In the case of GPS and CSS sources, the expansion speeds
$\sim 0.1 - 0.4 \> c$ of the radio plasma indicate that the filling
factor of dense clouds in the interstellar medium is relatively low
\citep{con02a,mur99a,par99,mur02a,mur03a,ows98,ows99}.  However, in
the case of our radio-excess objects, which show evidence of more
recent interaction or merging activity, the medium may not be as
clumped. This is consistent with the modest radio lobe expansion
velocities ($\lesssim 0.01 \> c$) inferred in \citet[hereafter
\paperthree]{dra04b}, and we comment further on this below.  We assume
that the optical line emission in the radio-excess objects is
predominantly shock-excited, and thus this model may be used to infer
the radio jet energy flux from the \oiiia\ luminosity. If there is a
significant contribution from photoionization or star formation, the
derived properties of the radio jets are overestimates.  The \oiiia\
luminosities and the jet energy fluxes, derived from Equation
\ref{spe_eqn:LOIII}, are listed in Table \ref{spe_tab:jet}.  The jet
energy fluxes are lower than for powerful radio sources, consistent
with the results of \paperthree, where we estimated jet energy fluxes
from the radio spectral energy distributions of several radio-excess
sources.  Objects with the highest radio powers also have the highest
\oiiia\ luminosities.  Thus the radio-loud objects (those with $u <
0$) have the highest jet energy fluxes, and objects with more moderate
radio excesses tend to have lower jet energy fluxes.

Let us examine these estimates for the energy flux in the light of the
velocity estimates of \paperthree. Using equation~(2.10) and (2.12) of
\citet{bic97}, the velocity of expansion of a radio lobe of size $x_h$
in a smooth medium with Hydrogen number density, $n_H$ is given by:
   \begin{eqnarray}
 \protect\vspace*{-3mm}  V_c \simeq 1500 \> \rm km \> s^{-1}  \, \left( \frac {6}{8-\delta}
\right)^{1/3}  \, \times \, & & \nonumber \\
  \protect\vspace*{3mm} \> \zeta^{1/6} \left( \frac {F_{E,45}}{ n_H({\rm kpc})} \right)^{1/3}  \,
   \left( \frac {x_h}{\rm kpc} \right)^{(\delta-2)/3} & & \nonumber \\
   \end{eqnarray}
where $\zeta \simeq 2$, and $F_{E,45}$ is the jet energy flux in units of
$10^{45}$~erg\,s$^{-1}$. Thus, the expansion velocities estimated in
\paperthree\ are consistent with $F_{E,45}/n_H({\rm kpc}) \sim 1$. In view of
our energy flux estimates, this implies that $n_H ({\rm kpc}) \sim 1$; for a
temperature $\sim 10^7$, the corresponding $p/k \sim 10^7$.

The energy fluxes in our radio-excess objects are much less than that in GPS
and CSS sources \citep{bic03a}, yet the [\ion{O}{3}] fluxes are
similar. Figure \ref{spe_fig:histOIIIrad} shows the distribution of \oiii\
luminosity to radio luminosity and it is clear that this parameter is
systematically higher for the radio excess objects compared to GPS and CSS
sources. This requires comment. There are two important effects to take into
account: (1) The lower expansion velocities of radio-excess objects mean that
it is easier for the shocks to be radiative in the sense that the radiative
timescale should be less than the dynamical time scale. (See \citealt{bic97}
and \citealt{bic03a} for discussions of this point.) (2) The covering factor
of dense clouds may be larger, consistent with our assumption of a medium with
a higher filling factor.

\subsection{Continuum Emission}
\label{spe_subsec:dis_con}

A number of objects show strong continuum as well as emission lines in
their spectra. In the case of a few, with broad emission lines, the
continuum is likely to be associated with the central nucleus. However
other objects show strong stellar continuum and absorption lines that
indicate the nature of the dominant stellar population in the host
galaxy.  A substantial fraction of the sample spectra show absorption
indicative of an intermediate age stellar population. Sixteen objects
(33\%) show \hbeta, or \hgamma, \hdelta, and \hepsilon\ absorption.
The stellar continuum and absorption in these objects are typical of A
type stars. Objects with this type of optical spectrum have previously
been described as post-starburst AGN \citep{dey94}. They are composite
objects that are believed to have undergone a starburst within the
past $\sim$ 10$^8$ years, and that are now dominated by an
intermediate-age stellar population and an active nucleus.  A further
13 objects in the sample (27\%) show redder stellar continuum and
absorption typical of elliptical galaxies.  These objects resemble
normal radio-loud objects for which the AGN host galaxies are luminous
elliptical galaxies.

\begin{figure}[H]
\epsscale{0.8}
\plotone{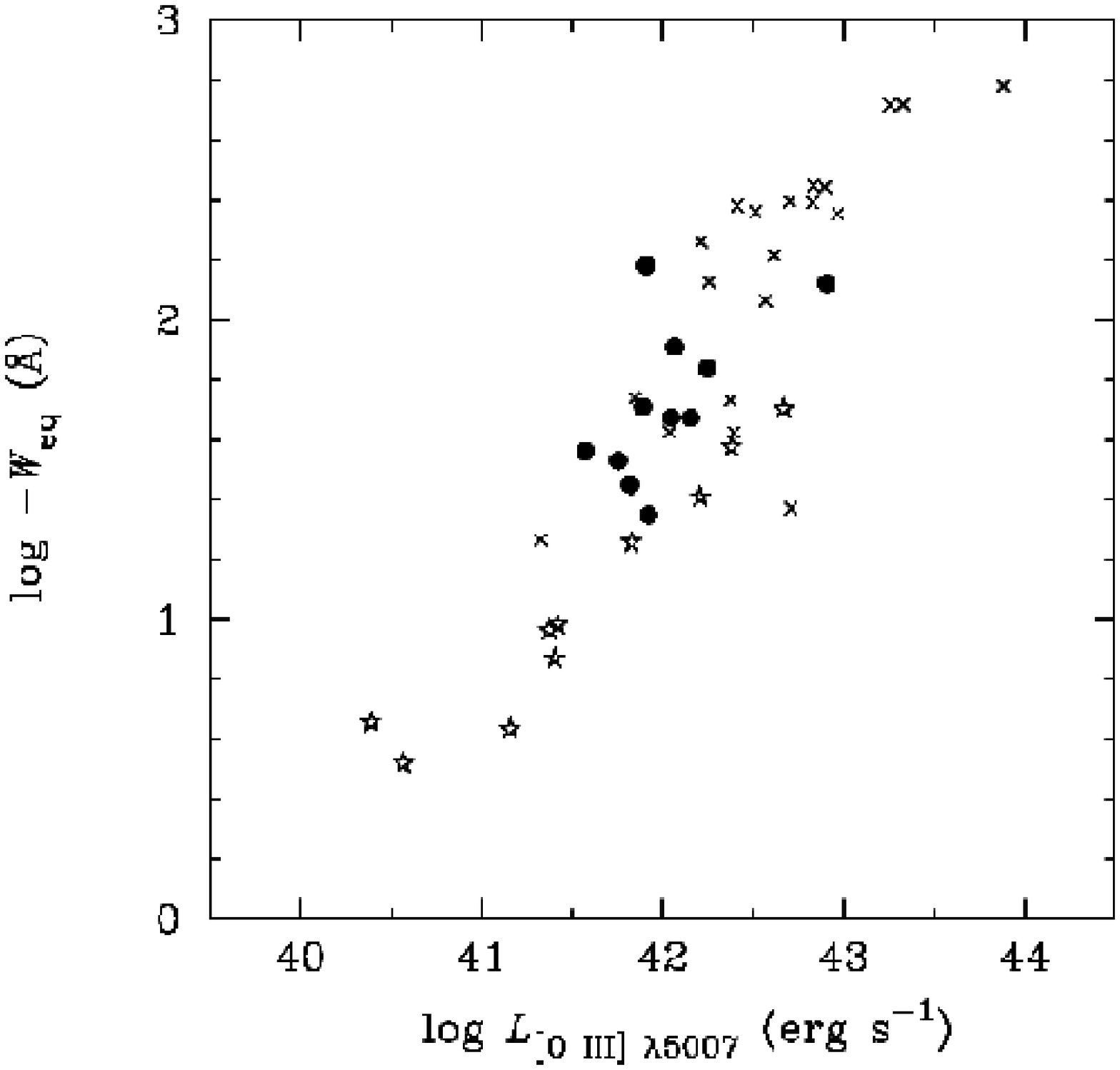}
\caption{\small \protect\oiiia\ equivalent widths versus \oiiia\ luminosity
as a function of classification by stellar continuum type. The symbols
indicate the type of stellar continuum in the spectrum; A-star spectra
(\emph{open stars}), older stellar continuum, typical of elliptical
galaxies (\emph{filled circles}), and objects with minimal stellar
continuum (\emph{x's}). \label{spe_fig:eqw_st}}
\epsscale{1.0}
\end{figure}

Figure \ref{spe_fig:eqw_st} shows the equivalent widths of the \oiiia\
emission lines as a function of \oiiia\ luminosity.  The symbols
indicate the type of stellar population, either A-star (\emph{open
stars}), or old elliptical stellar population (\emph{filled
circles}). Also plotted are objects showing minimal stellar continuum
in their spectrum (\emph{x's}), for which the nature of the
underlying continuum cannot be established. The equivalent widths and
luminosities of the \oiiia\ emission lines show a correlation because
the range of continuum luminosities in the sample is narrow compared
with the range of line luminosities.  Figure \ref{spe_fig:eqw_st}
shows that the objects with minimal stellar continuum all have high
equivalent widths and luminous \oiiia\ emission. In contrast, the
objects with A-star stellar continuum have low to moderate \oiiia\
equivalent widths and luminosities. Objects with an older stellar
population, typical of elliptical galaxies, have intermediate
equivalent widths and luminosities.

\begin{figure}[H]
\epsscale{0.8}
\plotone{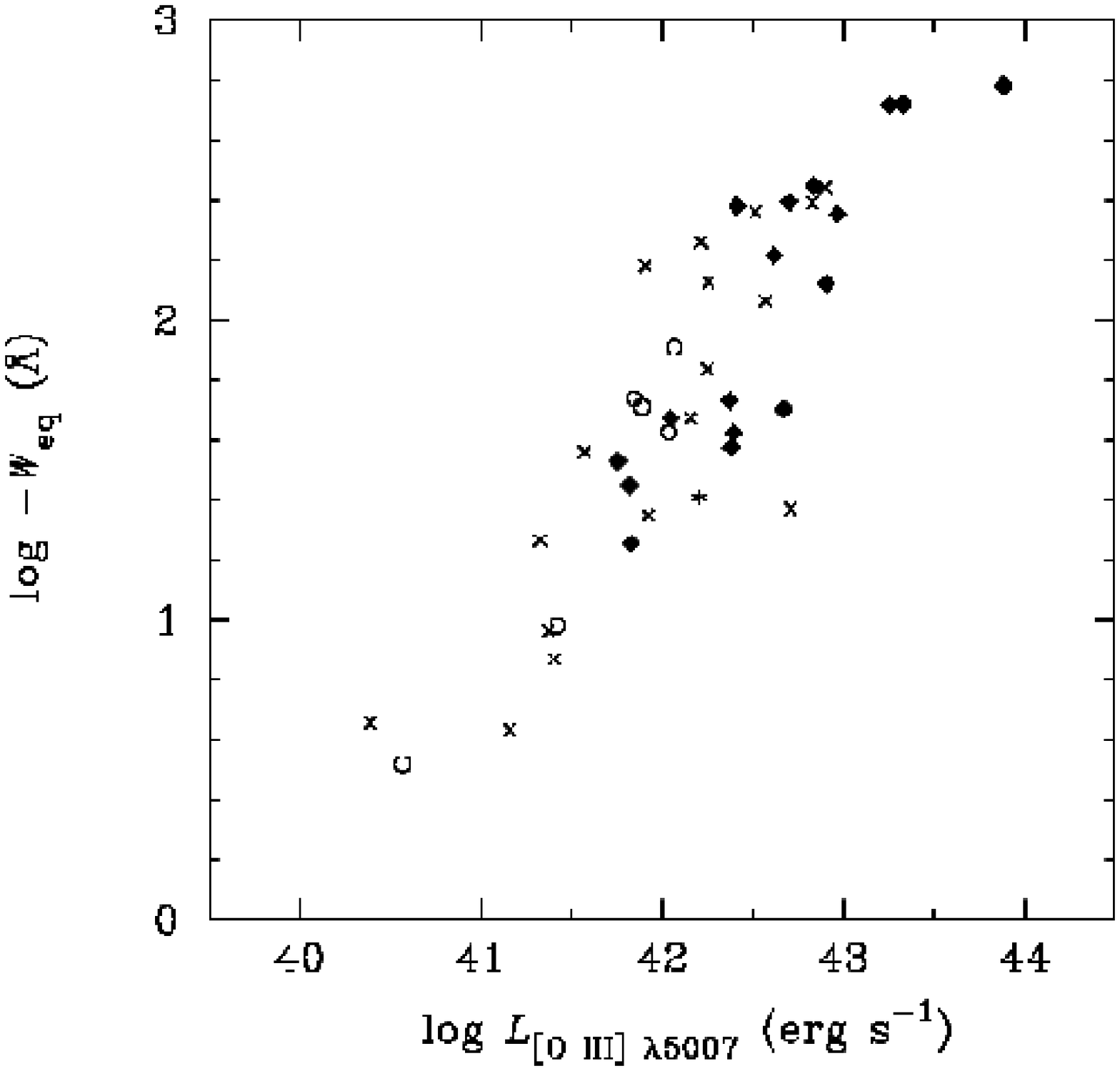}
\caption{\small \protect\oiiia\ equivalent widths versus \oiiia\ luminosity
as a function of morphological type for the FIR-luminous radio-excess
sample. Symbols indicate the host morphology, either disk (\emph{open
circles}), elliptical (\emph{filled squares}), point source
(\emph{stars}), or undetermined
(\emph{x's}). \label{spe_fig:eqw_ho}}
\epsscale{1.0}
\end{figure}

\begin{figure}[H]
\epsscale{0.8}
\plotone{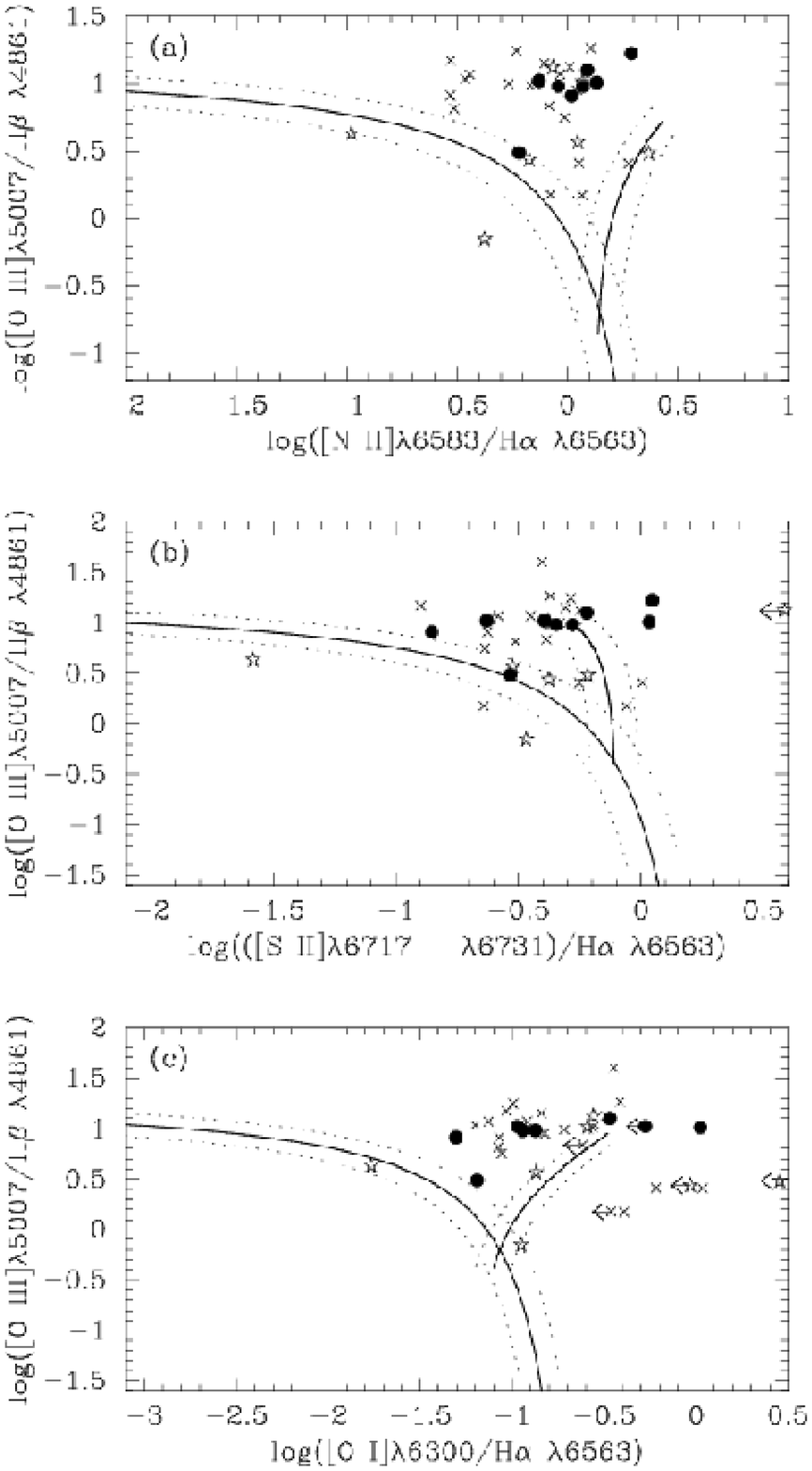}
\caption{\small Diagnostic diagrams. The symbols indicate the presence of a
  dominant post-starburst population (\emph{open stars}), old stellar
  population (\emph{filled circles}), or no stellar absorption in the
  spectrum (\emph{x's}).\label{spe_fig:diag_stellar}}
\epsscale{1.0}
\end{figure}

The distribution of points on this diagram may indicate that there are
different types of objects in the sample; post-starburst galaxies and
ellipticals with higher intrinsic luminosities. Alternatively, it may
reflect different ages of objects in the sample, indicating an aging
sequence from post-starburst galaxy to evolved elliptical to AGN. In
the first case, we would expect the different types of object to have
different host morphologies. In the second case, we would expect the
morphological type of the host galaxy to be unrelated to the location
of each object in Figure \ref{spe_fig:eqw_st}. We show in Figure
\ref{spe_fig:eqw_ho} the \oiiia\ equivalent widths and luminosities
plotted with different symbols indicating the host morphological type
determined in \papertwo. Morphological types determined by
surface-brightness fitting were disks (\emph{open circles}) and
ellipticals/bulges (\emph{filled squares}), point sources
(\emph{star}), and undetermined (\emph{x's}).  There is a clear
separation of the points, with disk hosts having low equivalent widths
and luminosities and elliptical hosts having high equivalent widths
and luminosities. This confirms the result found in \papertwo\ that
there are different types of object in the FIR-luminous radio-excess
sample. Taken together, Figures \ref{spe_fig:eqw_st} and
\ref{spe_fig:eqw_ho} suggest that disk hosts have post-starburst type
spectra, while elliptical hosts have an older stellar population or no
stellar continuum. However, we do not find a clear relation between
the continuum type and the host type. The stellar continuum type is
probably more fundamentally related to the weakness of the active
nucleus.  The objects that show strong stellar absorption also show
more dominant starburst activity in their optical spectra.  Figure
\ref{spe_fig:diag_stellar} shows the diagnostic diagrams, with the
same symbols as in Figure \ref{spe_fig:eqw_st} showing the objects
with A-star stellar continuum, older stellar continuum, and minimal
stellar continuum in their spectra. The post-starburst objects tend to
cluster around the maximum starburst and mixing lines and have
ambiguous or borderline classifications. The objects with older
stellar continuum or no stellar continuum tend to fall higher in the
diagnostic diagrams, indicating a more dominant AGN.

\begin{figure}[H]
\epsscale{0.8}
\plotone{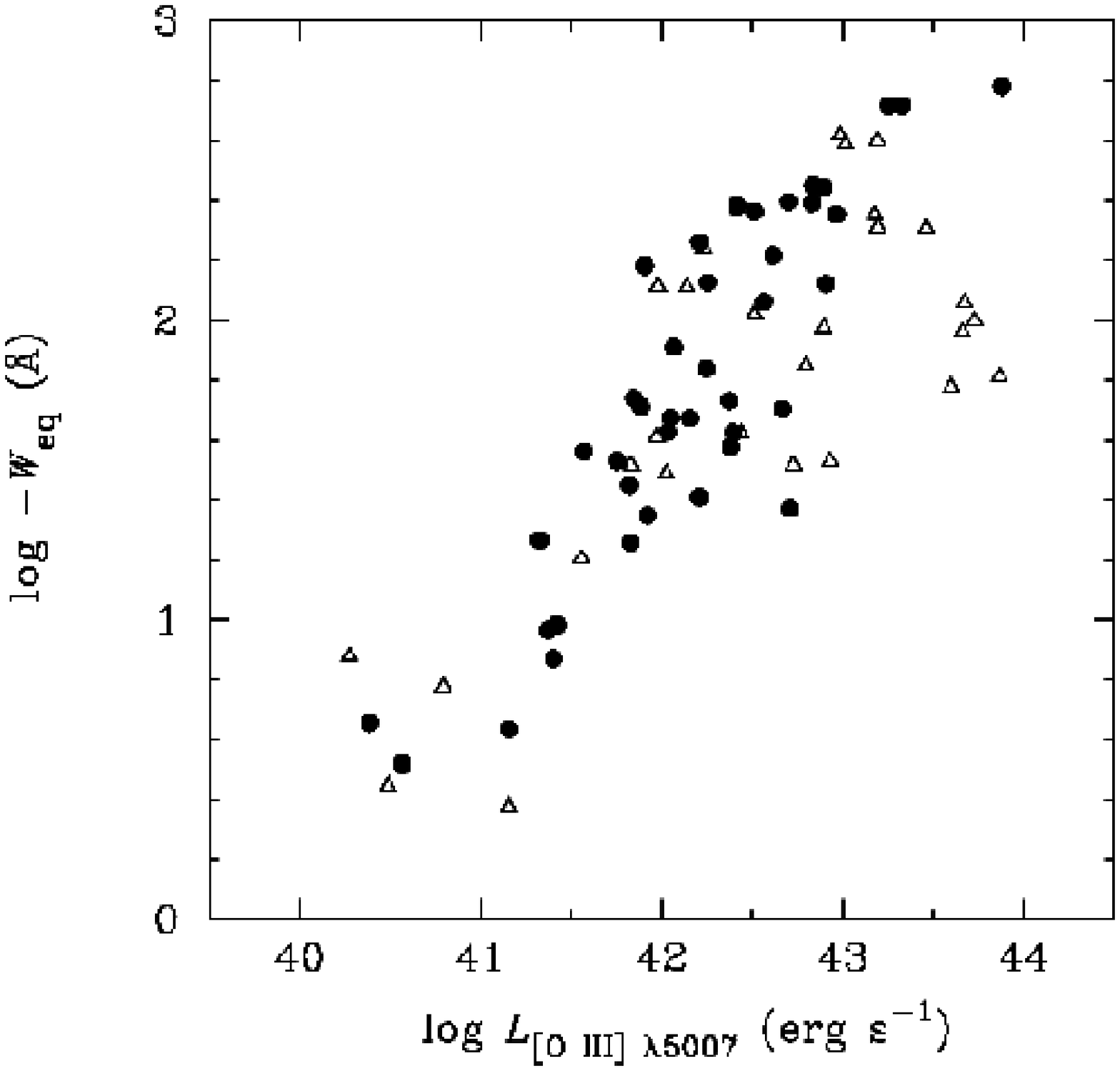}
\caption{\small \protect\oiiia\ equivalent widths versus \oiiia\ luminosity
for the FIR-luminous radio-excess sample (\emph{filled circles}) and
CSS sources (\emph{open triangles}; \protect\citealt{gel94}).
\label{spe_fig:eqw}} 
\epsscale{1.0}
\end{figure}

Figure \ref{spe_fig:eqw} shows the same equivalent width -- luminosity
diagram as Figures \ref{spe_fig:eqw_st} and \ref{spe_fig:eqw_ho}, with
data for CSS sources from \citet{gel94} plotted for comparison.  The
FIR-luminous radio-excess and CSS samples follow the same trend of
\oiiia\ equivalent width versus luminosity. We suggest that the four
CSS objects in the lower left part of the diagram are likely to be
post-starburst AGN. Examination of the spectra presented in
\citet{gel94} supports this hypothesis:  all of the four CSS sources
with \oiiia\ luminosities less than $10^{41.5}$ \ergs\ show Mg~b and
Na~D stellar absorption and two show \hbeta\ absorption, indicative of
an intermediate-age stellar population.

\subsection{The Nature of Radio-Excess \iras\ Galaxies}
\label{spe_subsec:dis_nat}

Radio-excess galaxies have been identified in small numbers by previous
authors but, prior to this work, have not previously been studied as a
class. Our multi-wavelength study has shown that these objects are more common
than generally believed and that the radio excess is an excellent indicator of
the presence of an active nucleus.  Most of the FIR-luminous radio-excess
objects have compact radio sources, as shown by our radio observations
(\paperthree). Sixty-five percent of the radio sources are less than 10 kpc in
size and have radio sizes an order of magnitude smaller than the optical
extent. The more extended objects are classical radio galaxies but the compact
radio sources appear to be a different population of objects.

The compact FIR-luminous radio-excess galaxies have steep radio spectra
indicative of optically thin synchrotron emission, but radio powers in between
those of traditional Seyfert and \cg\ sources.  Modeling of the radio data
suggests that the radio sources have ages 10$^4$ -- 10$^6$ years
(\paperthree), placing them among the young radio sources.  Synchrotron and
jet models used to interpret the radio and optical data show that the radio
jets in the radio-excess galaxies are intrinsically less powerful than radio
sources such as \cg\ sources.  This suggests that the intrinsic jet energy
flux is the dominant factor determining the weaker radio power in the
radio-intermediate sources.  We interpret the high ratios of \oiiia\
luminosity to radio power in the radio-excess objects to be a result of
slower, more luminous shocks in the radio-excess sources than in more powerful
radio sources.  This is supported by the derived expansion velocities ($< 0.01
c$), which are significantly less than those observed in \cg\ sources.  The
FIR-luminous radio-excess objects also show evidence for strong interaction
between the radio jet and the interstellar medium. Properties of the optical
emission lines indicate the effect of the jet and expanding radio lobe on the
kinematics of the line-emitting region.  Two representative radio-excess
sources observed with VLBI imaging, F11243-2244 and F23400-8705, have
kiloparsec-scale lobes. Their morphologies are not similar to the symmetric
doubles seen among the classes of \cg\ sources and suggest that the weaker
radio jets are disrupted by interaction with the interstellar medium.  The
compact radio-excess sources are not likely to evolve into classical radio
galaxies. Although we showed in \papertwo\ that the host galaxies are more
like those of \frii\ radio galaxies than \fris, the radio powers are expected
to decrease as the objects evolve, so that, if they were to expand to sizes of
hundreds of kiloparsecs, the radio-excess objects would be unlikely to have
\frii\ radio powers.

The radio-intermediate sources may be confined by the host interstellar
medium; further work is required to determine if this is likely. An \ion{H}{1}
absorption study would allow us to estimate the gas masses in the radio-excess
galaxies and determine if this is sufficient to confine the radio plasma.

The radio emission may be a transient phenomenon, perhaps associated with
additional fueling to the black hole accretion disk in the nucleus.  The
FIR-luminous radio-excess objects may then fade to become radio-quiet
galaxies.  This would be consistent with the large fraction of mergers in the
sample.  The optical morphologies of the host galaxies of the sample objects
include a large fraction of merging or disturbed galaxies (\papertwo).  The
post-starburst continuum spectra observed in one third of the sample provide
further evidence that many of the hosts have undergone interactions.  The
dynamical timescale for merging activity and the age of the A-star population
responsible for the post-starburst spectrum are both $\sim 10^{8}$ years. The
radio sources are at least two orders of magnitude younger than
this. Therefore, if the radio activity is associated with the tidal
interaction, then there must be a delay between the interaction and the
initiation of radio activity.

There may be some influence of the host type on the radio power; radio sources
with large radio excesses are not found in disk host galaxies
(\papertwo). Model-fitting of host surface brightness profiles suggests that
there are two populations of host galaxy in the FIR-luminous radio-excess
sample: disk and elliptical systems, with different host luminosities and
radio excess distributions (\papertwo). The disk systems have low radio
excesses and host luminosities, while the elliptical hosts have a broad range
of radio excesses and high optical luminosities.  This result suggests that an
elliptical host is necessary but not sufficient for a galaxy to host a
radio-loud nucleus.  However, the presence of both disk and elliptical
galaxies together with the presence of objects in our sample that lie between
radio-loud and radio-quiet objects in their radio/far-infrared and
radio/\oiiia\ luminosity ratios shows that the classic dichotomy between
radio--loud and radio--quiet galaxies, at the very least needs to be modified.
The compact radio-intermediate objects that appear in our sample are AGN with
relatively weak radio activity and a range of host morphologies, that fill the
gap between powerful CSS/GPS sources and radio-quiet Seyfert galaxies.

\section{Conclusions}
\label{spe_sec:con}

Analysis of the optical spectra of our sample of FIR-luminous radio-excess
galaxies confirms that radio-excess is an extraordinarily good indicator of
the presence of an AGN. We find a much higher fraction of AGN than samples
selected on FIR luminosity alone, or using other criteria such as warm FIR
colors. We find several ambiguous or borderline objects in the sample. These
are likely to be composite objects that comprise starbursts which also host an
AGN. There is some indication that the type of optical spectrum is associated
with the radio-loudness, in that the radio-loud objects appear to be more
`pure' AGN than radio-intermediate objects. They may be more powerful AGN,
though they still appear to be highly obscured.

In the \oiiia\ luminosity versus radio power diagram, the objects fall between
the known correlations observed for radio galaxies and Seyferts.  We interpret
this in the context of shock models describing the interaction of the jet with
the surrounding medium and infer that radio-intermediate objects generally
have lower jet powers than more powerful radio sources.  From our
understanding that the jet powers of radio-excess objects are intrinsically
lower than in \cg\ sources (\paperthree), we infer that the radio-excess
objects are equally luminous in \oiiia\ because the covering factor of dense
gas is larger.  This is supported by the lower expansion velocities inferred
for the radio-excess objects (\paperthree).

Despite the lower jet powers, the FIR-luminous radio-excess galaxies show
clear evidence for strong interaction between the radio jet and the
interstellar medium. Broad \oiiia\ line widths indicate acceleration of the
line-emitting material by the radio source. A blueshift is observed between
the \oiiia\ and \halpha\ emission lines in several sources. Line profiles of
two objects observed at higher spectral resolution show complex
structure. These further suggest that the kinematics of the line-emitting
material is strongly influenced by the radio source.

We find that 33\% of the sample spectra show post-starburst stellar
continuum, with A-star absorption lines. A further 13\% show old, red
stellar continuum and absorption typical of elliptical galaxies.  The
objects with post-starburst spectra appear to have weaker AGN; they
have low \oiiia\ luminosities and equivalent widths, and lie closer to
the starburst line in diagnostic diagrams. The objects with older
stellar continuum show intermediate \oiiia\ luminosities and
equivalent widths, and lie more clearly in the region of diagnostic
diagrams occupied by Seyfert galaxies.  Objects with minimal stellar
continuum have high luminosities and equivalent widths, and also
occupy the Seyfert region of the diagnostic diagrams.  The optical
emission from these objects is clearly dominated by the active
nucleus.

The presence of post-starburst AGN in the sample is consistent with
the large number of merging or disturbed host galaxies found in the
sample (\papertwo). If the radio sources are associated
with the merging activity, there must be a delay between the
interaction and the initiation of the radio activity. The A-star
spectra indicate ages of $\sim 10^{8}$ years for the dominant stellar
population, consistent with the dynamical timescales involved in
mergers \citep{bar96}. The ages of the radio sources are likely to be at least
two orders of magnitude younger than this.

Combining the understanding we have gleaned from our multi-wavelength study,
we find that the radio-intermediate objects span a range of properties between
those of powerful radio sources, such as \cg\ sources, and radio-quiet Seyfert
galaxies.  It remains unclear whether the radio-intermediate objects reveal a
continuum of radio-loudness in AGN or indicate that the distributions of
radio-loud and radio-quiet AGN overlap. The compact radio-intermediate objects
in the FIR-luminous radio-excess sample appear to be a new population of
relatively young radio sources that will not evolve into classical radio
galaxies, but which are more powerful than typical radio-quiet
Seyferts. Further observations are required to determine if the radio plasma
is confined by the host interstellar medium.

\acknowledgements

CLB wishes to acknowledge the support of an Australian Postgraduate Research
Award and a Duffield Scholarship.  MAD wishes to acknowledge the support of
the Australian National University and the Australian Research Council (ARC)
through his ARC Australian Federation Fellowship, and also under ARC Discovery
project DP0208445. GVB acknowledges support from the ARC Large Grant
A699050341.  We thank the referee for helpful comments which improved the
manuscript.

{}
\parskip 1.5ex
\normalsize

\onecolumn
\begin{deluxetable}{lrrrrrr}
\tabletypesize{\scriptsize}
\tablewidth{0pt}
\tablecolumns{7}
\tablecaption{\halpha\ and \hbeta\ fluxes and extinction.\label{spe_tab:lines_a}}
\tablehead{
\colhead{FSC Name} &
\colhead{$z$} &
\multicolumn{2}{c}{\halpha\ Flux \tablenotemark{a}} & 
\multicolumn{2}{c}{\hbeta\ Flux \tablenotemark{a}} & 
\colhead{E(B-V)}  \\ 
 &
 &
\colhead{narrow} &
\colhead{broad} &
\colhead{narrow} &
\colhead{broad} &
\\ 
}
\startdata
F00022+1231 & 0.143 &    15.80  $\pm$    0.4 & \nodata\phn\phn\phn   &     4.11  $\pm$    0.2 & \nodata\phn\phn\phn   & \phs  0.2 \\ 
F00183-7111 & 0.327 &     5.11  $\pm$    0.3 & \nodata\phn\phn\phn   &     1.74  $\pm$    0.6 & \nodata\phn\phn\phn   & $>$  0.0 \\ 
F01009-3241 & 0.256 &    13.80  $\pm$    0.3 & \nodata\phn\phn\phn   &     0.70  $\pm$    0.1 & \nodata\phn\phn\phn   & \phs  1.8 \\ 
F01477-1655 & 0.162 &    54.50  $\pm$    1.5 & \nodata\phn\phn\phn   &    16.04  $\pm$    1.0 & \nodata\phn\phn\phn   & \phs  0.1 \\ 
F02483-5456 & 0.394 &     3.57  $\pm$    0.2 & \nodata\phn\phn\phn   &     0.40  $\pm$    0.2 & \nodata\phn\phn\phn   & \phs  1.1 \\ 
F03169+1850 & 0.300 &     1.41  $\pm$    0.1 & \nodata\phn\phn\phn   & $<$    0.94\phs\phn\phn\phn\phn   & \nodata\phn\phn\phn   & $>$  0.0 \\ 
F03265-2852 & 0.108 &    29.10  $\pm$    0.9 & \nodata\phn\phn\phn   &     8.61  $\pm$    0.7 & \nodata\phn\phn\phn   & \phs  0.1 \\ 
F04137-1144 & 0.090 &     3.93  $\pm$    0.8 & \nodata\phn\phn\phn   & $<$    6.73\phs\phn\phn\phn\phn   & \nodata\phn\phn\phn   & $>$  0.0 \\ 
F04367-2726 & 0.111 &    27.40  $\pm$    0.7 & \nodata\phn\phn\phn   &     4.51  $\pm$    0.5 & \nodata\phn\phn\phn   & \phs  0.7 \\ 
F05246+0103 & 0.097 &    17.10  $\pm$    0.8 & \nodata\phn\phn\phn   &     3.12  $\pm$    0.6 & \nodata\phn\phn\phn   & \phs  0.6 \\ 
F05265-4720 & 0.134 &     3.88  $\pm$    0.3 & \nodata\phn\phn\phn   & $<$    3.92\phs\phn\phn\phn\phn   & \nodata\phn\phn\phn   & $>$  0.0 \\ 
F06065-7934 & 0.287 &     4.97  $\pm$    1.3 & \nodata\phn\phn\phn   & $<$    0.93\phs\phn\phn\phn\phn   & \nodata\phn\phn\phn   & $>$  0.5 \\ 
F06195-4503 & 0.206 &     7.38  $\pm$    0.3 & \nodata\phn\phn\phn   &     1.22  $\pm$    0.2 & \nodata\phn\phn\phn   & \phs  0.7 \\ 
F08064-1018 & 0.110 &    39.50  $\pm$    0.8 & \nodata\phn\phn\phn   &    10.22  $\pm$    0.4 & \nodata\phn\phn\phn   & \phs  0.2 \\ 
F09165-0726 & 0.273 &     1.09  $\pm$    0.1 & \nodata\phn\phn\phn   & $<$    1.32\phs\phn\phn\phn\phn   & \nodata\phn\phn\phn   & $>$  0.0 \\ 
F09323+0732 & 0.290 &     5.30  $\pm$    0.2 & \nodata\phn\phn\phn   &     1.55  $\pm$    0.2 & \nodata\phn\phn\phn   & \phs  0.1 \\ 
F09416-0137 & 0.112 &    11.70  $\pm$    0.8 & \nodata\phn\phn\phn   &     0.55  $\pm$    0.5 & \nodata\phn\phn\phn   & \phs  1.9 \\ 
F10227-8251 & 0.086 &    11.40  $\pm$    0.4 & \nodata\phn\phn\phn   &     2.30  $\pm$    0.3 & \nodata\phn\phn\phn   & \phs  0.5 \\ 
F10418+1153 & 0.230 &     2.69  $\pm$    0.1 & \nodata\phn\phn\phn   & $<$    1.97\phs\phn\phn\phn\phn   & \nodata\phn\phn\phn   & $>$  0.0 \\ 
F11243-2244 & 0.171 &    12.20  $\pm$    0.2 &    20.80  $\pm$    0.5 &     0.89  $\pm$    0.1 &     0.64  $\pm$    0.4 & \phs  1.5 \\ 
F11270+0031 & 0.211 &     2.34  $\pm$    0.1 & \nodata\phn\phn\phn   &     0.77  $\pm$    0.1 & \nodata\phn\phn\phn   & $>$  0.0 \\ 
F12173-3541 & 0.057 &     5.93  $\pm$    0.2 & \nodata\phn\phn\phn   &     0.54  $\pm$    0.1 & \nodata\phn\phn\phn   & \phs  1.3 \\ 
F12183-1015 & 0.300 &     0.91  $\pm$    0.1 & \nodata\phn\phn\phn   & $<$    0.51\phs\phn\phn\phn\phn   & \nodata\phn\phn\phn   & $>$  0.0 \\ 
F14381-3901 & 0.138 &    43.10  $\pm$    0.8 &    26.20  $\pm$    4.4 &     9.14  $\pm$    0.4 &     5.18  $\pm$    3.2 & \phs  0.4 \\ 
F14501+0639 & 0.267 &     6.13  $\pm$    0.3 & \nodata\phn\phn\phn   & $<$    4.12\phs\phn\phn\phn\phn   & \nodata\phn\phn\phn   & $>$  0.0 \\ 
F15129+0432 & 0.095 &     7.13  $\pm$    0.2 & \nodata\phn\phn\phn   &     1.20  $\pm$    0.1 & \nodata\phn\phn\phn   & \phs  0.7 \\ 
F15494-7905 & 0.150 &     9.07  $\pm$    0.3 &    15.30  $\pm$    1.0 &     0.66  $\pm$    0.2 &     2.01  $\pm$    0.8 & \phs  1.5 \\ 
F15599+0206 & 0.104 &    20.90  $\pm$    0.5 & \nodata\phn\phn\phn   &     5.18  $\pm$    0.3 & \nodata\phn\phn\phn   & \phs  0.3 \\ 
F16159-0402 & 0.213 &    21.70  $\pm$    0.4 & \nodata\phn\phn\phn   &     3.11  $\pm$    0.4 & \nodata\phn\phn\phn   & \phs  0.8 \\ 
F16431+0217 & 0.096 &    20.90  $\pm$    0.6 & \nodata\phn\phn\phn   &     4.21  $\pm$    0.5 & \nodata\phn\phn\phn   & \phs  0.5 \\ 
F17345+1124 & 0.165 &    56.60  $\pm$    0.9 & \nodata\phn\phn\phn   &    14.21  $\pm$    0.6 & \nodata\phn\phn\phn   & \phs  0.3 \\ 
F19254-7245 & 0.062 &    18.40  $\pm$    0.5 &   130.00  $\pm$    2.0 &     1.88  $\pm$    0.3 &     1.37  $\pm$    1.5 & \phs  1.2 \\ 
F20203-5733 & 0.352 &    29.70  $\pm$    0.5 &   120.00  $\pm$    3.0 &     8.12  $\pm$    0.3 &    10.60  $\pm$    2.2 & \phs  0.2 \\ 
F20517-5809 & 0.193 &     4.53  $\pm$    0.4 & \nodata\phn\phn\phn   & $<$    2.85\phs\phn\phn\phn\phn   & \nodata\phn\phn\phn   & $>$  0.0 \\ 
F21139-6613 & 0.153 &    71.50  $\pm$    1.2 & \nodata\phn\phn\phn   &     1.02  $\pm$    0.3 & \nodata\phn\phn\phn   & \phs  3.1 \\ 
F21309-0726 & 0.087 &    12.00  $\pm$    0.3 & \nodata\phn\phn\phn   &     1.42  $\pm$    0.2 & \nodata\phn\phn\phn   & \phs  1.0 \\ 
F21356-1015 & 0.206 &    18.50  $\pm$    0.4 &    17.70  $\pm$    1.0 &     1.97  $\pm$    0.2 &     1.35  $\pm$    0.7 & \phs  1.1 \\ 
F21497-0824 & 0.035 &    73.80  $\pm$    1.2 & \nodata\phn\phn\phn   &    13.45  $\pm$    0.6 & \nodata\phn\phn\phn   & \phs  0.6 \\ 
F21511-4606 & 0.146 &     9.67  $\pm$    0.3 &    14.10  $\pm$    2.6 &     1.79  $\pm$    0.2 &     2.38  $\pm$    1.9 & \phs  0.6 \\ 
F21527-2856 & 0.232 & $<$    1.95\phs\phn\phn\phn\phn   & \nodata\phn\phn\phn   & $<$    0.94\phs\phn\phn\phn\phn   & \nodata\phn\phn\phn   & \nodata \\ 
F22537-6512 & 0.120 &    45.80  $\pm$    1.0 & \nodata\phn\phn\phn   &    11.65  $\pm$    0.6 & \nodata\phn\phn\phn   & \phs  0.2 \\ 
F23002-5828 & 0.210 &    26.50  $\pm$    1.5 & \nodata\phn\phn\phn   & $<$   11.11\phs\phn\phn\phn\phn   & \nodata\phn\phn\phn   & $>$  0.0 \\ 
F23075-5957 & 0.141 &     1.48  $\pm$    0.4 & \nodata\phn\phn\phn   &     0.95  $\pm$    0.3 & \nodata\phn\phn\phn   & $>$  0.0 \\ 
F23140+0348 & 0.220 &    12.90  $\pm$    0.4 & \nodata\phn\phn\phn   &     0.86  $\pm$    0.2 & \nodata\phn\phn\phn   & \phs  1.6 \\ 
F23389+0300 & 0.145 &    15.10  $\pm$    0.3 & \nodata\phn\phn\phn   &     2.63  $\pm$    0.3 & \nodata\phn\phn\phn   & \phs  0.6 \\ 
F23400-8705 & 0.106 &    12.10  $\pm$    0.4 & \nodata\phn\phn\phn   & $<$    7.18\phs\phn\phn\phn\phn   & \nodata\phn\phn\phn   & $>$  0.0 \\ 
F23475-7407 & 0.105 & $<$    6.02\phs\phn\phn\phn\phn   & \nodata\phn\phn\phn   & $<$    4.78\phs\phn\phn\phn\phn   & \nodata\phn\phn\phn   & \nodata \\ 
F23493-0126 & 0.174 &    25.40  $\pm$    1.1 &   327.00  $\pm$    9.7 &     6.24  $\pm$    0.8 &    71.10  $\pm$    7.1 & \phs  0.3 \\ 
F23565-7631 & 0.082 &     0.93  $\pm$    0.4 & \nodata\phn\phn\phn   &     0.36  $\pm$    0.3 & \nodata\phn\phn\phn   & $>$  0.0 \\ 
\enddata
\tablenotetext{a}{Fluxes are in units of $10^{-15}$ \ergsc.\\ \halpha\ and \hbeta\ fluxes have not been corrected for reddening.}
\end{deluxetable}

\begin{deluxetable}{lrrrrrrrrrrrr}
\tabletypesize{\scriptsize}
\tablewidth{0pt}
\tablecolumns{17}
\tablecaption{Emission line fluxes.\label{spe_tab:lines_b}}
\rotate
\tablehead{
\colhead{FSC Name} &
\colhead{[Ne$\:$\textsc{V}]} &
\colhead{[O$\:$\textsc{II}]} &
\colhead{[Ne$\:$\textsc{III}]} &
\colhead{[O$\:$\textsc{III}]} &
\colhead{He$\:$\textsc{II}} &
\colhead{[O$\:$\textsc{III}]} &
\colhead{[O$\:$\textsc{III}]} &
\colhead{[O$\:$\textsc{I}]} &
\colhead{[N$\:$\textsc{II}]} &
\colhead{[N$\:$\textsc{II}]} &
\colhead{[S$\:$\textsc{II}]} &
\colhead{[S$\:$\textsc{II}]} \\ 
  &
\colhead{$\lambda$3426} &
\colhead{$\lambda$3727} &
\colhead{$\lambda$3869} &
\colhead{$\lambda$4363} &
\colhead{$\lambda$4686} &
\colhead{$\lambda$4959} &
\colhead{$\lambda$5007} &
\colhead{$\lambda$6300} &
\colhead{$\lambda$6548} &
\colhead{$\lambda$6584} &
\colhead{$\lambda$6717} &
\colhead{$\lambda$6731} 
 }
\startdata
F00022+1231 &     4.72 &    15.99 &     5.10 &     1.08 &     2.02 &    17.14 &    51.06 &     2.03 &     2.53 &     7.44 &     3.11 &     4.35  \\ 
F00183-7111 & $<$    1.18 &     6.85 &     1.73 & $<$    1.48 & $<$    1.49 &     1.62 &     4.49 &     3.08 &     1.93 &     5.77 &     3.19 &     1.98  \\ 
F01009-3241 &   703.98 &  4763.\phn\phn   &  1267.\phn\phn   & $<$  511.11 & $<$  323.88 &   873.13 &  3262.\phn\phn   &   211.35 &   271.00 &   695.68 &   109.18 &   124.70  \\ 
F01477-1655 &    16.22 &    41.94 &    18.22 &     3.62 &     5.49 &    77.52 &   228.76 &     4.12 &     7.69 &    22.57 &     8.78 &     7.03  \\ 
F02483-5456 & $<$  106.25 &   119.89 & $<$   82.46 & $<$   53.68 & $<$   44.55 &    36.95 &   125.51 &     8.11 &     8.72 &    25.24 & $<$   56.84 & $<$   56.48  \\ 
F03169+1850 & $<$    0.33 &     0.73 &     0.53 & $<$    0.52 &     0.33 & $<$    2.26 &     2.54 & $<$    2.10 &     0.91 &     2.92 &     0.41 &     0.08  \\ 
F03265-2852 &    19.14 &    25.69 &    11.15 & $<$    7.22 &     5.70 &    35.52 &   100.10 &     5.22 &    13.08 &    38.24 &    11.10 &     7.83  \\ 
F04137-1144 & $<$    2.15 &     0.83 & $<$    1.25 & $<$    1.16 & $<$    1.23 & $<$    3.82 & $<$    3.64 & $<$    4.38 &     0.76 &     2.24 &     0.80 &     0.69  \\ 
F04367-2726 &    20.75 &   110.59 & $<$   74.33 & $<$   58.76 & $<$   38.08 &    42.13 &   104.34 &     6.77 &    21.84 &    63.55 &    17.40 &    13.36  \\ 
F05246+0103 & $<$   38.67 & $<$   35.00 &    49.74 & $<$   28.12 & $<$   22.38 & $<$   31.88 &    25.98 & $<$   18.48 &    15.38 &    44.80 &     8.22 &     4.02  \\ 
F05265-4720 & $<$    1.41 &     4.20 &     1.44 & $<$    2.00 & $<$    2.24 & $<$    2.86 & $<$    2.86 & $<$    2.90 &     1.53 &     4.50 &     1.22 &     0.56  \\ 
F06065-7934 & $<$    4.40 &    10.18 &     6.86 & $<$    5.07 & $<$    4.66 &     6.53 &    16.95 & $<$    4.49 &     5.30 &    15.48 &     0.29 &     4.97  \\ 
F06195-4503 &    12.09 &    25.66 &    22.68 & $<$   10.24 & $<$   11.93 &    19.45 &    62.04 & $<$    6.77 &     7.96 &    23.21 &     6.89 &     4.68  \\ 
F08064-1018 &    26.75 &    40.28 &    20.71 &     4.97 &     7.98 &    76.02 &   232.52 &     4.56 &     7.57 &    22.15 &     9.48 &     6.55  \\ 
F09165-0726 & $<$    0.66 &     1.09 & $<$    0.81 & $<$    1.12 & $<$    1.02 &     0.52 &     0.70 & $<$    1.54 &     0.32 &     0.93 &     0.38 &     0.35  \\ 
F09323+0732 &     1.07 &     7.08 &     3.17 & $<$    2.51 & $<$    2.12 &     7.52 &    26.28 &     2.20 &     2.71 &     7.95 &     2.14 &     1.74  \\ 
F09416-0137 & $<$ 7219.\phn\phn   &  6818.\phn\phn   & 13400.\phn\phn   &   723.02 & $<$ 2167.\phn\phn   &   447.74 &   644.93 &    73.23 &   223.86 &   606.02 &    86.26 &    76.79  \\ 
F10227-8251 & $<$   26.64 &    52.81 &    14.61 & $<$   21.82 & $<$   21.96 &    32.69 &    90.10 &     3.30 &     9.04 &    26.37 &     8.25 &     7.02  \\ 
F10418+1153 & $<$    1.06 &     3.05 &     1.99 &     0.59 & $<$    1.64 &     2.62 &     6.41 &     0.47 &     0.97 &     3.19 &     0.76 &     0.61  \\ 
F11243-2244 &   449.08 &   544.20 &   166.28 & $<$  169.48 &    15.57 &   186.32 &   626.47 &    11.79 &    86.09 &   248.39 &    15.49 &    17.90  \\ 
F11270+0031 &     0.84 &     2.49 &     0.89 &     0.30 &     0.28 &     2.65 &     7.74 &     0.64 &     0.43 &     1.27 &     0.62 &     0.47  \\ 
F12173-3541 & $<$  929.58 &   416.77 & $<$  428.82 &    23.67 & $<$  198.12 &    42.48 &    65.99 & $<$   69.23 &    17.36 &    50.09 &    22.45 &     9.03  \\ 
F12183-1015 & $<$    0.31 &     1.01 & $<$    0.37 & $<$    0.42 &     0.38 & $<$    0.66 & $<$    0.68 &     0.24 &     0.24 &     0.72 &     0.27 &     0.17  \\ 
F14381-3901 & $<$   15.53 &   103.49 &    29.89 &    12.64 &     3.50 &    87.52 &   262.43 &     8.45 &    10.05 &    29.45 &    14.65 &     8.94  \\ 
F14501+0639 & $<$    0.73 &     1.62 &     0.75 & $<$    1.73 & $<$    1.89 &     1.55 &     6.97 &     1.11 &     1.56 &     4.60 &     0.46 &     1.42  \\ 
F15129+0432 & $<$   52.36 &    48.62 & $<$   42.56 &     2.05 & $<$   23.21 &     2.84 &     5.95 &     2.92 &     3.79 &    11.03 &     5.58 &     3.32  \\ 
F15494-7905 & $<$  771.10 &  1068.\phn\phn   &   538.72 & $<$  258.17 & $<$  215.70 &   925.16 &  2271.\phn\phn   &    62.59 &    75.23 &   216.39 &    39.80 &    30.02  \\ 
F15599+0206 &    24.10 &    31.21 &    13.95 &     6.59 &     5.26 &    54.16 &   161.66 &     5.11 &     9.50 &    27.88 &    10.92 &     6.53  \\ 
F16159-0402 & $<$   51.40 &   288.32 &    35.13 & $<$   33.07 & $<$   32.20 &    21.74 &    52.83 &    44.38 &    44.08 &   128.18 &    13.06 &    82.66  \\ 
F16431+0217 &    24.42 &    42.67 &    14.20 &    12.66 &     8.37 &    54.23 &   165.90 &     7.18 &    21.60 &    62.99 &    12.37 &    11.74  \\ 
F17345+1124 &     7.68 &    77.84 &    23.09 &    22.08 &     8.11 &    99.00 &   292.63 &    17.91 &    21.63 &    63.32 &    20.59 &    19.59  \\ 
F19254-7245 & $<$  975.59 &   963.08 & $<$  478.67 & $<$  383.22 & $<$  287.96 &   251.71 &   602.16 &   192.02 &    85.52 &   247.36 &   103.95 &    93.19  \\ 
F20203-5733 &    13.31 &    29.71 &    16.05 &     4.13 &     3.41 &    65.30 &   198.66 &     3.84 &     4.18 &    12.27 &     2.94 &     2.28  \\ 
F20517-5809 & $<$    1.89 &     2.45 & $<$    1.94 & $<$    2.52 &     1.67 & $<$    3.49 & $<$    3.49 &     0.58 &     1.17 &     3.45 &     0.80 &     1.04  \\ 
F21139-6613 & $<$695400.\phn\phn   & 583700.\phn\phn   & 92600.\phn\phn   & 16400.\phn\phn   & $<$78500.\phn\phn   & 21900.\phn\phn   & 50000.\phn\phn   &   615.32 &  1353.\phn\phn   &  3796.\phn\phn   &   505.94 &   440.09  \\ 
F21309-0726 &   203.05 &   158.70 &    88.64 & $<$  112.09 &    29.65 &   115.27 &   299.30 & $<$   46.76 &    22.58 &    65.48 &    12.47 &     8.29  \\ 
F21356-1015 & $<$  215.96 &   752.80 &   184.77 &    24.91 & $<$   78.89 &   374.18 &   960.54 &    17.09 &    34.40 &    99.46 &    87.49 &     0.26  \\ 
F21497-0824 & $<$  102.02 &   194.04 &    94.31 &    20.42 &    26.21 &   265.04 &   783.17 &    24.66 &    58.92 &   172.12 &    49.04 &    44.07  \\ 
F21511-4606 &    25.95 &    82.71 &    26.73 &    11.58 &     3.23 &    47.10 &   125.63 &     9.42 &    10.31 &    30.02 &     8.97 &     7.54  \\ 
F21527-2856 & \nodata & \nodata & \nodata & \nodata & \nodata & \nodata & \nodata & \nodata & \nodata & \nodata & \nodata & \nodata  \\ 
F22537-6512 &    41.74 &    75.74 &    34.18 &     4.71 &    13.37 &    96.19 &   280.10 &     8.89 &    23.34 &    68.32 &    13.51 &    12.59  \\ 
F23002-5828 & $<$    3.21 &     4.96 & $<$    3.67 & $<$    3.46 & $<$    3.22 & $<$    4.86 &     1.18 & $<$    5.84 &     0.36 &     1.06 &     1.59 &     0.59  \\ 
F23075-5957 & $<$    3.48 &     1.93 & $<$    4.53 & $<$    4.27 & $<$    4.85 &     2.29 &     2.89 & $<$    4.21 &     1.18 &     3.46 &     0.52 &     0.38  \\ 
F23140+0348 &   549.98 &  5150.\phn\phn   &   363.80 & $<$  853.04 & $<$  612.95 &   480.54 &   997.58 &    76.25 &   121.01 &   347.55 &    64.23 &    60.54  \\ 
F23389+0300 & $<$   21.91 &   109.67 &     8.55 & $<$   18.01 & $<$   14.28 &    19.72 &    42.62 &    55.90 &    33.43 &    97.38 &    15.57 &    13.22  \\ 
F23400-8705 & $<$    2.96 &     3.31 & $<$    3.84 & $<$    3.85 & $<$    4.28 &     4.88 &    13.05 &     1.26 &     3.35 &     9.86 &     2.75 &     2.04  \\ 
F23475-7407 & \nodata & \nodata & \nodata & \nodata & \nodata & \nodata & \nodata & \nodata & \nodata & \nodata & \nodata & \nodata  \\ 
F23493-0126 &    11.50 &    30.58 &    11.32 &    43.38 & $<$   42.89 &    26.52 &    79.56 &     3.84 &    14.51 &    42.43 &     5.06 &     5.07  \\ 
F23565-7631 & $<$    2.50 &     1.45 & $<$    2.66 & $<$    3.60 & $<$    3.92 & $<$    6.19 & $<$    5.99 & $<$    6.70 &     0.62 &     1.81 &     0.29 &     0.74  \\ 
\enddata
\tablecomments{Fluxes are in units of $10^{-15}$ \ergsc.\\ All fluxes have been corrected for reddening, listed in Table \ref{spe_tab:lines_a}.}
\end{deluxetable}

\begin{deluxetable}{lcccccccccccccccc}
\tabletypesize{\scriptsize}
\tablewidth{0pt}
\tablecolumns{18}
\tablecaption{Emission line widths.\label{spe_tab:fwhm}}
\rotate
\tablehead{
\colhead{FSC Name} &
\colhead{[Ne$\:$\textsc{V}]} &
\colhead{[O$\:$\textsc{II}]} &
\colhead{[Ne$\:$\textsc{III}]} &
\colhead{[O$\:$\textsc{III}]} &
\colhead{He$\:$\textsc{II}} &
\multicolumn{2}{c}{\hbeta} & 
\colhead{[O$\:$\textsc{III}]} &
\colhead{[O$\:$\textsc{III}]} &
\colhead{[O$\:$\textsc{I}]} &
\multicolumn{2}{c}{\halpha} & 
\colhead{[N$\:$\textsc{II}]} &
\colhead{[N$\:$\textsc{II}]} &
\colhead{[S$\:$\textsc{II}]} &
\colhead{[S$\:$\textsc{II}]} \\ 
  &
\colhead{$\lambda$3426} &
\colhead{$\lambda$3727} &
\colhead{$\lambda$3869} &
\colhead{$\lambda$4363} &
\colhead{$\lambda$4686} &
\multicolumn{2}{c}{narrow broad} & 
\colhead{$\lambda$4959} &
\colhead{$\lambda$5007} &
\colhead{$\lambda$6300} &
\multicolumn{2}{c}{narrow broad} & 
\colhead{$\lambda$6548} &
\colhead{$\lambda$6584} &
\colhead{$\lambda$6717} &
\colhead{$\lambda$6731} 
 }
\startdata
F00022+1231 &   279  &   395  &   522  &   364  &   266  &   439  & \nodata  &   343 &   342  &   720  &   476  & \nodata  &   376  &   374  &   403  &   651 \\ 
F00183-7111 & \nodata  &   871  &  1375  & \nodata  & \nodata  &  2272  & \nodata  &   795 &   893  &   563  &   720  & \nodata  &   950  &   944  &   942  &   940 \\ 
F01009-3241 &   419  &   688  &   725  & \nodata  & \nodata  &   511  & \nodata  &   633 &   887  &   599  &   770  & \nodata  &   769  &   679  &   683  &   681 \\ 
F01477-1655 &   419  &   513  &   408  &   588  &   439  &   542  & \nodata  &   443 &   451  &   329  &   573  & \nodata  &   459  &   457  &   422  &   366 \\ 
F02483-5456 & \nodata  &   751  & \nodata  & \nodata  & \nodata  &   762  & \nodata  &   914 &   974  &   323  &   691  & \nodata  &   693  &   689  & \nodata  & \nodata \\ 
F03169+1850 & \nodata  &   327  &  1107  & \nodata  &   187  & \nodata  & \nodata  & \nodata &  1408  & \nodata  &   635  & \nodata  &   979  &  1063  &  1030  &  1027 \\ 
F03265-2852 &   818  &   419  &   563  & \nodata  &   572  &   506  & \nodata  &   542 &   496  &   555  &   467  & \nodata  &   516  &   513  &   455  &   412 \\ 
F04137-1144 & \nodata  &   606  & \nodata  & \nodata  & \nodata  & \nodata  & \nodata  & \nodata & \nodata  & \nodata  &  1642  & \nodata  &   430  &   428  &   471  &   335 \\ 
F04367-2726 &   590  &   372  & \nodata  & \nodata  & \nodata  &   500  & \nodata  &   513 &   425  &   624  &   506  & \nodata  &   532  &   529  &   558  &   557 \\ 
F05246+0103 & \nodata  & \nodata  &  1040  & \nodata  & \nodata  &   207  & \nodata  & \nodata &   479  & \nodata  &   175  & \nodata  &   471  &   468  &   262  &   214 \\ 
F05265-4720 & \nodata  &   484  &   775  & \nodata  & \nodata  & \nodata  & \nodata  & \nodata & \nodata  & \nodata  &   678  & \nodata  &   364  &   362  &   366  &   366 \\ 
F06065-7934 & \nodata  &   481  &   780  & \nodata  & \nodata  & \nodata  & \nodata  &  1671 &  1335  & \nodata  &   506  & \nodata  &   638  &   634  &  1058  &  1055 \\ 
F06195-4503 &   516  &   592  &  1061  & \nodata  & \nodata  &   406  & \nodata  &   256 &   573  & \nodata  &   514  & \nodata  &   515  &   512  &   469  &   468 \\ 
F08064-1018 &   435  &   452  &   482  &   728  &   381  &   486  & \nodata  &   467 &   472  &   511  &   459  & \nodata  &   482  &   479  &   452  &   422 \\ 
F09165-0726 & \nodata  &   538  & \nodata  & \nodata  & \nodata  & \nodata  & \nodata  &   723 &   951  & \nodata  &   572  & \nodata  &   523  &   521  &  1030  &  1028 \\ 
F09323+0732 &   663  &   515  &   768  & \nodata  & \nodata  &   827  & \nodata  &   731 &   978  &   697  &   493  & \nodata  &   756  &   752  &   470  &   381 \\ 
F09416-0137 & \nodata  &   460  &  2377  &  1027  & \nodata  &   384  & \nodata  &  1478 &  1095  &   510  &   524  & \nodata  &   599  &   559  &   529  &   528 \\ 
F10227-8251 & \nodata  &   605  &   662  & \nodata  & \nodata  &   429  & \nodata  &   427 &   395  &   415  &   399  & \nodata  &   398  &   396  &   366  &   418 \\ 
F10418+1153 & \nodata  &   505  &   807  &   681  & \nodata  & \nodata  & \nodata  &   836 &   560  &   155  &   216  & \nodata  &   225  &   296  &   249  &   249 \\ 
F11243-2244 &  1590  &   563  &   348  & \nodata  &   164  & 664  & 2439  &   688 &   759  &   304  & 695  & 2448  &   697  &   693  &   698  &   697 \\ 
F11270+0031 &   590  &   517  &   518  &    72  &   356  &   382  & \nodata  &   358 &   293  &   532  &   471  & \nodata  &   320  &   319  &   325  &   325 \\ 
F12173-3541 & \nodata  &   511  & \nodata  &   613  & \nodata  &   480  & \nodata  &  1826 &  1104  & \nodata  &   355  & \nodata  &   399  &   397  &   532  &   272 \\ 
F12183-1015 & \nodata  &   528  & \nodata  & \nodata  &   184  & \nodata  & \nodata  & \nodata & \nodata  &   116  &   372  & \nodata  &   394  &   392  &   384  &   384 \\ 
F14381-3901 & \nodata  &   598  &   572  &  1217  &   376  & 656  & 7234  &   556 &   558  &   568  & 482  & 7238  &   481  &   478  &   538  &   364 \\ 
F14501+0639 & \nodata  &   570  &  1108  & \nodata  & \nodata  & \nodata  & \nodata  &   623 &   836  &   810  &   790  & \nodata  &   432  &   430  &   691  &   690 \\ 
F15129+0432 & \nodata  &   511  & \nodata  &   526  & \nodata  &   275  & \nodata  &   117 &   411  &   326  &   204  & \nodata  &   259  &   257  &   260  &   218 \\ 
F15494-7905 & \nodata  &   649  &  1025  & \nodata  & \nodata  & 665  & 2297  &  1676 &  1439  &   974  & 696  & 2306  &   697  &   693  &   458  &   577 \\ 
F15599+0206 &   511  &   562  &   516  &   888  &   599  &   575  & \nodata  &   504 &   489  &   493  &   484  & \nodata  &   488  &   486  &   542  &   388 \\ 
F16159-0402 & \nodata  &  1016  &   872  & \nodata  & \nodata  &  1335  & \nodata  &  1572 &  1510  &   915  &   771  & \nodata  &   779  &   775  &  1223  &  1221 \\ 
F16431+0217 &   633  &   615  &   431  &  1329  &   802  &   693  & \nodata  &   642 &   646  &   605  &   514  & \nodata  &   622  &   618  &   648  &   646 \\ 
F17345+1124 &   412  &   918  &   709  &  3195  &  1387  &  1025  & \nodata  &   922 &   897  &  1002  &   873  & \nodata  &   772  &   767  &   808  &   807 \\ 
F19254-7245 & \nodata  &   812  & \nodata  & \nodata  & \nodata  & 459  & 2565  &  1823 &  1578  &  1650  & 524  & 2578  &   525  &   522  &   562  &   561 \\ 
F20203-5733 &   453  &   608  &   493  &   575  &   674  & 718  & 8079  &   576 &   477  &   708  & 628  & 8084  &   630  &   626  &   374  &   373 \\ 
F20517-5809 & \nodata  &   973  & \nodata  & \nodata  &   447  & \nodata  & \nodata  & \nodata & \nodata  &   164  &   627  & \nodata  &   859  &   855  &   356  &   411 \\ 
F21139-6613 & \nodata  &   735  &   516  &   412  & \nodata  &   609  & \nodata  &   508 &   735  &   369  &   482  & \nodata  &   461  &   459  &   432  &   399 \\ 
F21309-0726 &   493  &   340  &   680  & \nodata  &   709  &   336  & \nodata  &   737 &   579  & \nodata  &   399  & \nodata  &   368  &   366  &   375  &   255 \\ 
F21356-1015 & \nodata  &   759  &   882  &   584  & \nodata  & 505  & 2701  &  1455 &  1470  &   585  & 570  & 2714  &   571  &   568  &  1254  &  1304 \\ 
F21497-0824 & \nodata  &   589  &   722  &   776  &   760  &   655  & \nodata  &   731 &   723  &   698  &   594  & \nodata  &   582  &   579  &   481  &   505 \\ 
F21511-4606 &   471  &   455  &   429  &   444  &   284  & 357  & 4969  &   432 &   368  &   387  & 472  & 4974  &   577  &   574  &   392  &   294 \\ 
F21527-2856 & \nodata  &   716  & \nodata  & \nodata  & \nodata  & \nodata  & \nodata  & \nodata &   625  & \nodata  & \nodata  & \nodata  & \nodata  & \nodata  & \nodata  & \nodata \\ 
F22537-6512 &   658  &   606  &   655  &   672  &   617  &   617  & \nodata  &   593 &   575  &   515  &   440  & \nodata  &   516  &   514  &   397  &   397 \\ 
F23002-5828 & \nodata  &   711  & \nodata  & \nodata  & \nodata  & \nodata  & \nodata  & \nodata &   979  & \nodata  &  1955  & \nodata  &   677  &   673  &   678  &   677 \\ 
F23075-5957 & \nodata  &   814  & \nodata  & \nodata  & \nodata  &   388  & \nodata  &   101 &   554  & \nodata  &   567  & \nodata  &   459  &   456  &   315  &   282 \\ 
F23140+0348 &   449  &  1018  &   484  & \nodata  & \nodata  &   721  & \nodata  &  1399 &  1386  &   897  &   769  & \nodata  &   770  &   766  &   453  &   452 \\ 
F23389+0300 & \nodata  &   803  &   549  & \nodata  & \nodata  &  1137  & \nodata  &  1864 &  1308  &  1988  &   729  & \nodata  &  1181  &  1174  &   517  &   516 \\ 
F23400-8705 & \nodata  &   406  & \nodata  & \nodata  & \nodata  & \nodata  & \nodata  &  1356 &  1476  &   803  &   558  & \nodata  &   525  &   523  &   562  &   561 \\ 
F23475-7407 & \nodata  & \nodata  & \nodata  & \nodata  & \nodata  & \nodata  & \nodata  & \nodata & \nodata  & \nodata  & \nodata  & \nodata  & \nodata  & \nodata  & \nodata  & \nodata \\ 
F23493-0126 &   567  &   724  &   770  &  4501  & \nodata  & 602  & 6563  &   639 &   607  &   459  & 622  & 6571  &   624  &   621  &   348  &   347 \\ 
F23565-7631 & \nodata  &   425  & \nodata  & \nodata  & \nodata  &   571  & \nodata  & \nodata & \nodata  & \nodata  &   629  & \nodata  &   466  &   463  &   341  &   340 \\ 
\enddata
\tablecomments{All linewidths are in \kms.}
\end{deluxetable}

\begin{deluxetable}{lrcclr}
\tabletypesize{\scriptsize}
\tablewidth{0pt}
\tablecolumns{2}
\tablecaption{\protect\oiiia\ equivalent widths.\label{spe_tab:EQ}}
\tablehead{
\colhead{FSC Name} &
\colhead{$W_{\rm{eq}}$} & 
 & &
\colhead{FSC Name} &
\colhead{$W_{\rm{eq}}$} \\ 
 &
\colhead{(\AA)} 
 & & & &
\colhead{(\AA)} 
}
\startdata
F00022+1231 & -241.70  & & &   F15129+0432 & -3.32 \\ 	      
F00183-7111 & -54.11   & & &   F15494-7905 & -231.60 \\      
F01009-3241 & -250.20  & & &   F15599+0206 & -116.00 \\      
F01477-1655 & -526.20  & & &   F16159-0402 & -42.42 \\       
F02483-5456 & -50.71   & & &   F16431+0217 & -69.19 \\       
F03169+1850 & -47.41   & & &   F17345+1124 & -524.30 \\      
F03265-2852 & -165.20  & & &   F19254-7245 & -36.60 \\       
F04137-1144 & -9.64    & & &   F20203-5733 & -607.80 \\      
F04367-2726 & -22.42   & & &   F20517-5809 & \nodata\phn \\  
F05246+0103 & -18.54   & & &   F21139-6613 & -18.13 \\       
F05265-4720 & -4.71    & & &   F21309-0726 & -34.06 \\       
F06065-7934 & -47.27   & & &   F21356-1015 & -278.90 \\      
F06195-4503 & -134.70  & & &   F21497-0824 & -152.60 \\      
F08064-1018 & -247.50  & & &   F21511-4606 & -42.28 \\       
F09165-0726 & -7.46    & & &   F21527-2856 & -9.25 \\ 	      
F09323+0732 & -133.30  & & &   F22537-6512 & -226.50 \\      
F09416-0137 & -4.32    & & &   F23002-5828 & \nodata\phn \\  
F10227-8251 & -51.60   & & &   F23075-5957 & -9.64 \\ 	      
F10418+1153 & -25.80   & & &   F23140+0348 & \nodata\phn \\  
F11243-2244 & -81.47   & & &   F23389+0300 & -55.05 \\       
F11270+0031 & -183.40  & & &   F23400-8705 & -28.28 \\       
F12173-3541 & -4.55    & & &   F23475-7407 & -0.84 \\ 	      
F12183-1015 & -3.34    & & &   F23493-0126 & -23.60 \\       
F14381-3901 & -282.00  & & &   F23565-7631 & -1.51 \\        
F14501+0639 & -37.89   & & &               &        \\ 
\enddata
\end{deluxetable}

\begin{deluxetable}{lcccc}
\tabletypesize{\scriptsize}
\tablewidth{0pt}
\tablecolumns{5}
\tablecaption{Object classifications.\label{spe_tab:class}}
\tablehead{
\colhead{FSC Name} &
\multicolumn{3}{c}{Individual classifications} & 
\colhead{Type} \\ 
  &
\colhead{\ratioNIIHa} &
\colhead{\ratioSIIHa} &
\colhead{\ratioOIHa} &
\colhead{} 
 }
\startdata
F00022+1231 & Sy & Sy & Sy & Sy \\ 
F00183-7111 & Sy & Sy & LN & AMB \\ 
F01009-3241 & Sy & Sy & Sy & Sy \\ 
F01477-1655 & Sy & Sy & Sy & Sy \\ 
F02483-5456 & Sy & (Sy) & Sy & Sy \\ 
F03169+1850 & \nodata & \nodata & \nodata & \nodata \\ 
F03265-2852 & Sy & Sy & Sy & Sy \\ 
F04137-1144 & \nodata & \nodata & \nodata & \nodata \\ 
F04367-2726 & Sy & Sy & Sy & Sy \\ 
F05246+0103 & Sy & SB & (LN) & AMB \\ 
F05265-4720 & \nodata & \nodata & \nodata & \nodata \\ 
F06065-7934 & \nodata & \nodata & \nodata & \nodata \\ 
F06195-4503 & Sy & Sy & (Sy) & Sy \\ 
F08064-1018 & Sy & Sy & Sy & Sy \\ 
F09165-0726 & \nodata & \nodata & \nodata & \nodata \\ 
F09323+0732 & Sy & Sy & Sy & Sy \\ 
F09416-0137 & Sy & Sy & Sy & Sy \\ 
F10227-8251 & Sy & Sy & Sy & Sy \\ 
F10418+1153 & \nodata & \nodata & \nodata & \nodata \\ 
F11243-2244 & Sy & Sy & Sy & Sy \\ 
F11270+0031 & Sy & Sy & Sy & Sy \\ 
F12173-3541 & Sy & Sy & (LN) & Sy \\ 
F12183-1015 & \nodata & \nodata & \nodata & \nodata \\ 
F14381-3901 & Sy & Sy & Sy & Sy \\ 
F14501+0639 & \nodata & \nodata & \nodata & \nodata \\ 
F15129+0432 & SB & SB & LN & AMB \\ 
F15494-7905 & Sy & Sy & Sy & Sy \\ 
F15599+0206 & Sy & Sy & Sy & Sy \\ 
F16159-0402 & Sy & Sy & LN & AMB \\ 
F16431+0217 & Sy & Sy & Sy & Sy \\ 
F17345+1124 & Sy & Sy & Sy & Sy \\ 
F19254-7245 & Sy & Sy & LN & AMB \\ 
F20203-5733 & Sy & Sy & Sy & Sy \\ 
F20517-5809 & \nodata & \nodata & \nodata & \nodata \\ 
F21139-6613 & SB & SB & SB & SB \\ 
F21309-0726 & Sy & Sy & (LN) & Sy \\ 
F21356-1015 & Sy & Sy & Sy & Sy \\ 
F21497-0824 & Sy & Sy & Sy & Sy \\ 
F21511-4606 & Sy & Sy & Sy & Sy \\ 
F21527-2856 & \nodata & \nodata & \nodata & \nodata \\ 
F22537-6512 & Sy & Sy & Sy & Sy \\ 
F23002-5828 & \nodata & \nodata & \nodata & \nodata \\ 
F23075-5957 & LN & Sy & (LN) & AMB \\ 
F23140+0348 & Sy & Sy & Sy & Sy \\ 
F23389+0300 & Sy & Sy & LN & AMB \\ 
F23400-8705 & \nodata & \nodata & \nodata & \nodata \\ 
F23475-7407 & \nodata & \nodata & \nodata & \nodata \\ 
F23493-0126 & Sy & Sy & Sy & Sy \\ 
F23565-7631 & (Sy) & (Sy) & (LN) & Sy \\ 
\enddata
\tablecomments{Parentheses indicate a classification made using a limit in the line ratio; these were not used in assigning the final type. Objects for which \halpha\ or \hbeta\ is a limit were not classified.}
\end{deluxetable}

\begin{deluxetable}{lcccc}
\tabletypesize{\scriptsize}
\tablewidth{0pt}
\tablecolumns{5}
\tablecaption{Jet energy fluxes.\label{spe_tab:jet}}
\tablehead{
\colhead{FSC Name} &
\colhead{$u$} &
\colhead{$\log L_{\:\rm{[O\:III]}}$} &
\colhead{$\log P_{\:\rm{5\:GHz}}$} &
\colhead{$\log F_E$} \\
 & &
\colhead{(W)} &
\colhead{(\whz)} &
\colhead{(W)}
}
\startdata
F00022+1231 &  -0.18 & \phs\phd 35.41 &  25.83 & \phs\phd 37.45  \\ 
F00183-7111 &   1.04 & \phs\phd 35.37 &  25.76 & \phs\phd 37.41  \\ 
F01009-3241 &   0.48 & \phs\phd 35.70 &  25.48 & \phs\phd 37.74  \\ 
F01477-1655 &   0.74 & \phs\phd 36.33 &  24.78 & \phs\phd 38.36  \\ 
F02483-5456 &   0.20 & \phs\phd 35.67 &  25.93 & \phs\phd 37.70  \\ 
F03169+1850 &  -0.45 & \phs\phd 35.05 &  26.50 & \phs\phd 37.08  \\ 
F03265-2852 &   0.03 & \phs\phd 35.61 &  25.36 & \phs\phd 37.65  \\ 
F04137-1144 &   0.86 & $<$ 34.12 &  24.42 & $<$ 36.16 \\ 
F04367-2726 &   1.11 & \phs\phd 34.92 &  24.40 & \phs\phd 36.96  \\ 
F05246+0103 &   1.15 & \phs\phd 34.33 &  24.88 & \phs\phd 36.36  \\ 
F05265-4720 &   1.27 & $<$ 34.37 &  24.60 & $<$ 36.41  \\ 
F06065-7934 &  -0.19 & \phs\phd 35.16 &  26.31 & \phs\phd 37.19  \\ 
F06195-4503 &  -0.24 & \phs\phd 35.25 &  25.99 & \phs\phd 37.29  \\ 
F08064-1018 &  -0.51 & \phs\phd 35.83 &  25.98 & \phs\phd 37.86  \\ 
F09165-0726 &   0.23 & \phs\phd 34.40 &  25.75 & \phs\phd 36.44  \\ 
F09323+0732 &  -0.13 & \phs\phd 35.91 &  26.16 & \phs\phd 37.94  \\ 
F09416-0137 &   1.22 & \phs\phd 34.16 &  24.60 & \phs\phd 36.19  \\ 
F10227-8251 &   1.07 & \phs\phd 34.89 &  24.02 & \phs\phd 36.92  \\ 
F10418+1153 &   0.44 & \phs\phd 35.21 &  25.38 & \phs\phd 37.24  \\ 
F11243-2244 &   0.67 & \phs\phd 35.07 &  24.94 & \phs\phd 37.10  \\ 
F11270+0031 &  -0.18 & \phs\phd 35.21 &  25.85 & \phs\phd 37.25  \\ 
F12173-3541 &   1.28 & \phs\phd 33.39 &  24.23 & \phs\phd 35.42  \\ 
F12183-1015 &   0.64 & $<$ 34.47 &  25.39 & $<$ 36.51  \\ 
F14381-3901 &   0.31 & \phs\phd 35.83 &  25.18 & \phs\phd 37.87  \\ 
F14501+0639 &   0.12 & \phs\phd 35.38 &  25.89 & \phs\phd 37.41  \\ 
F15129+0432 &   1.05 & \phs\phd 33.57 &  24.45 & \phs\phd 35.60  \\ 
F15494-7905 &  -0.56 & \phs\phd 35.51 &  26.58 & \phs\phd 37.55  \\ 
F15599+0206 &  -0.67 & \phs\phd 35.57 &  26.14 & \phs\phd 37.60  \\ 
F16159-0402 &   1.17 & \phs\phd 35.04 &  25.15 & \phs\phd 37.07  \\ 
F16431+0217 &  -0.42 & \phs\phd 35.25 &  25.43 & \phs\phd 37.28  \\ 
F17345+1124 &   1.14 & \phs\phd 36.25 &  24.65 & \phs\phd 38.29  \\ 
F19254-7245 &   1.76 & \phs\phd 34.57 &  24.21 & \phs\phd 36.61  \\ 
F20203-5733 &  -0.44 & \phs\phd 36.88 &  26.67 & \phs\phd 38.92  \\ 
F20517-5809 &   0.89 & $<$ 34.79 &  24.90 & $<$ 36.82  \\ 
F21139-6613 &   1.18 & \phs\phd 34.83 &  24.36 & \phs\phd 36.86  \\ 
F21309-0726 &   0.72 & \phs\phd 34.75 &  24.33 & \phs\phd 36.79  \\ 
F21356-1015 &   1.00 & \phs\phd 35.90 &  24.97 & \phs\phd 37.93  \\ 
F21497-0824 &   1.68 & \phs\phd 34.91 &  23.54 & \phs\phd 36.94  \\ 
F21511-4606 &   0.66 & \phs\phd 35.40 &  24.61 & \phs\phd 37.43  \\ 
F21527-2856 &   0.20 & \phs\phd 34.37 &  25.57 & \phs\phd 36.41  \\ 
F22537-6512 &   0.81 & \phs\phd 35.97 &  25.11 & \phs\phd 38.00  \\ 
F23002-5828 &   1.12 & \phs\phd 34.39 &  24.65 & \phs\phd 36.43  \\ 
F23075-5957 &   0.66 & \phs\phd 34.42 &  25.15 & \phs\phd 36.46  \\ 
F23140+0348 &  -0.35 & \phs\phd 35.39 &  26.49 & \phs\phd 37.43  \\ 
F23389+0300 &   0.80 & \phs\phd 34.84 &  25.28 & \phs\phd 36.88  \\ 
F23400-8705 &   0.32 & \phs\phd 34.82 &  24.73 & \phs\phd 36.86  \\ 
F23475-7407 &   1.08 & $<$ 34.40 &  23.86 & $<$ 36.43  \\ 
F23493-0126 &  -0.46 & \phs\phd 35.71 &  25.99 & \phs\phd 37.74  \\ 
F23565-7631 &   1.13 & $<$ 34.26 &  23.89 & $<$ 36.29  \\ 
\enddata
\end{deluxetable}

\twocolumn

\appendix

\section{Optical Spectra of FIR-Luminous Radio-Excess Objects} \label{spe_sec:app_spec}

Here we show the low-resolution optical spectra discussed in Section
\ref{spe_sec:obs} (Fig. \ref{spe_fig:spec}).
\begin{figure}
\protect\vspace*{-12in}
\epsscale{0.7}
\plotone{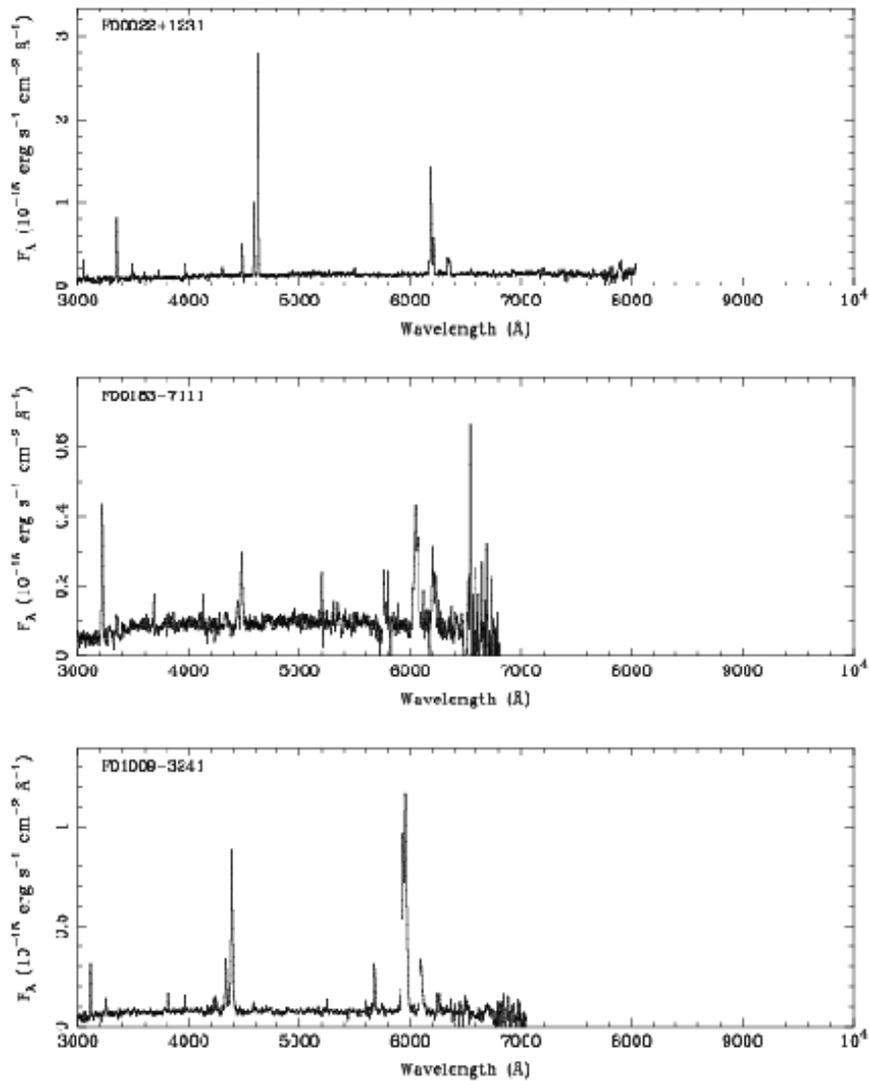}
\caption{\small Optical spectra of the FIR-luminous radio-excess \iras\
  galaxies. Spectra have been converted to restframe. The vertical
  axis is $F_{\lambda}$ in units of \ergscA.\label{spe_fig:spec}}
\epsscale{1.0}
\end{figure}

\begin{figure}[H]
\addtocounter{figure}{-1}
\epsscale{0.7}
\plotone{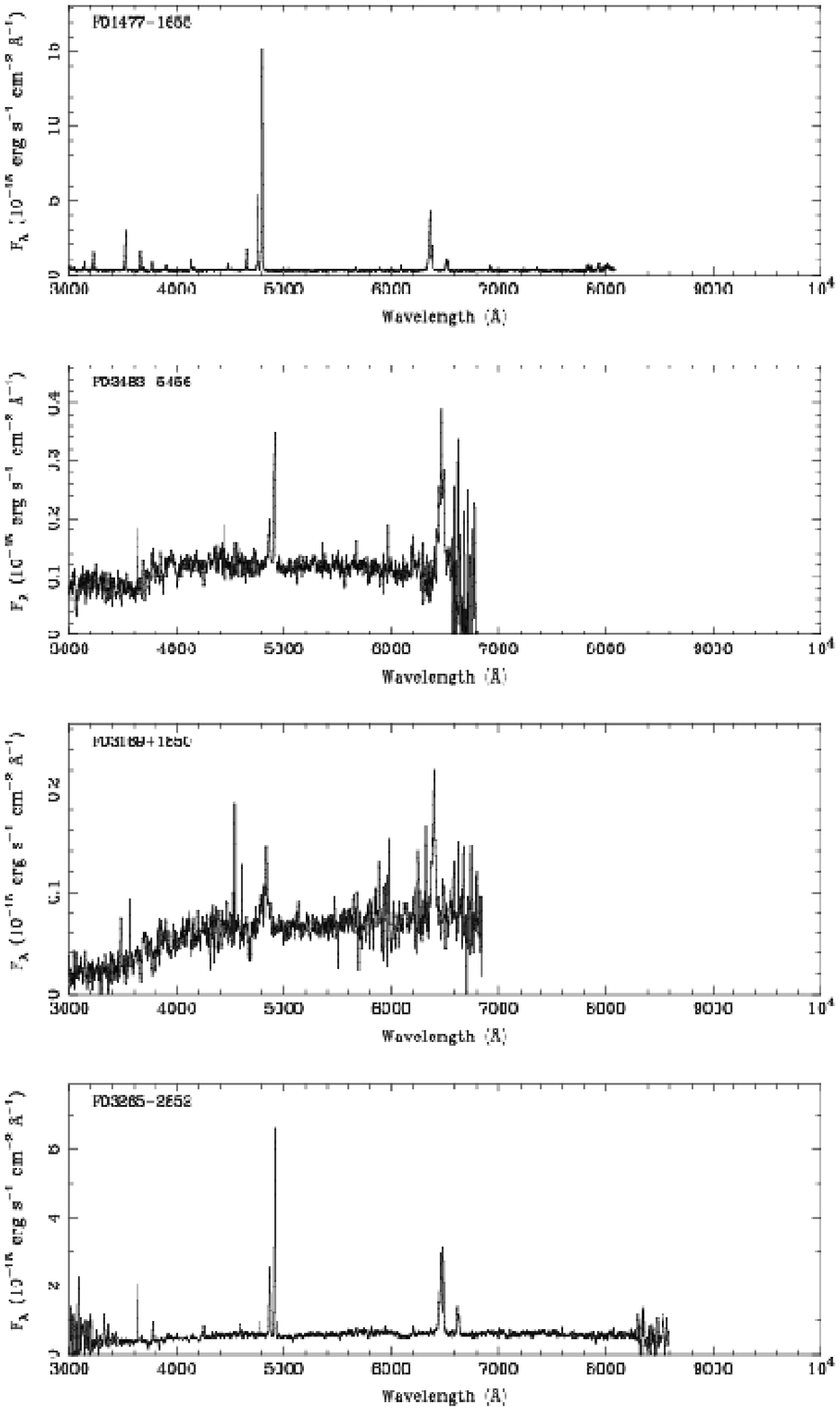}
\caption{\small \protect\emph{Continued.}}
\epsscale{1.0}
\end{figure}

\begin{figure}[H]
\addtocounter{figure}{-1}
\epsscale{0.7}
\plotone{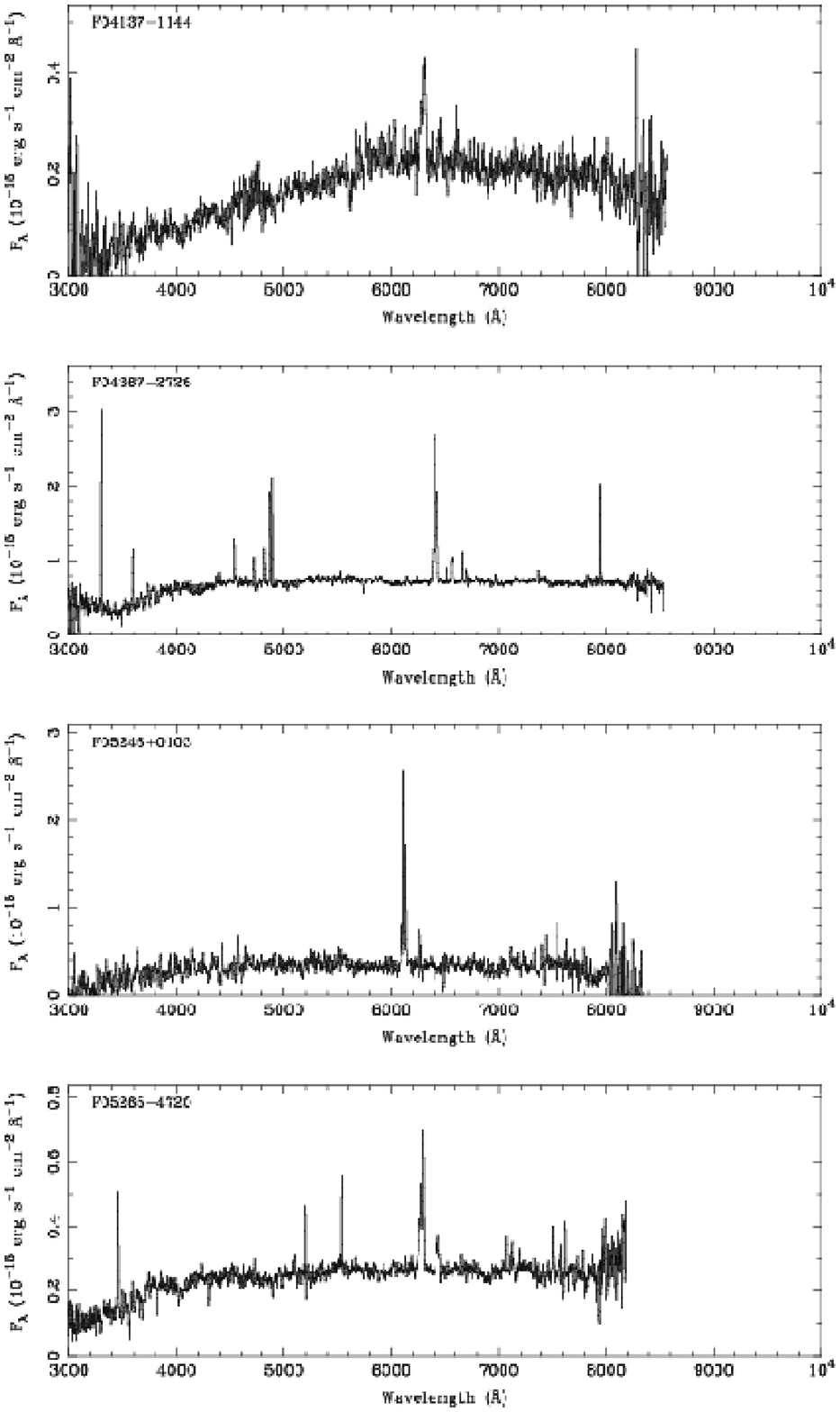}
\caption{\small \protect\emph{Continued.}}
\epsscale{1.0}
\end{figure}

\begin{figure}[H]
\addtocounter{figure}{-1}
\epsscale{0.7}
\plotone{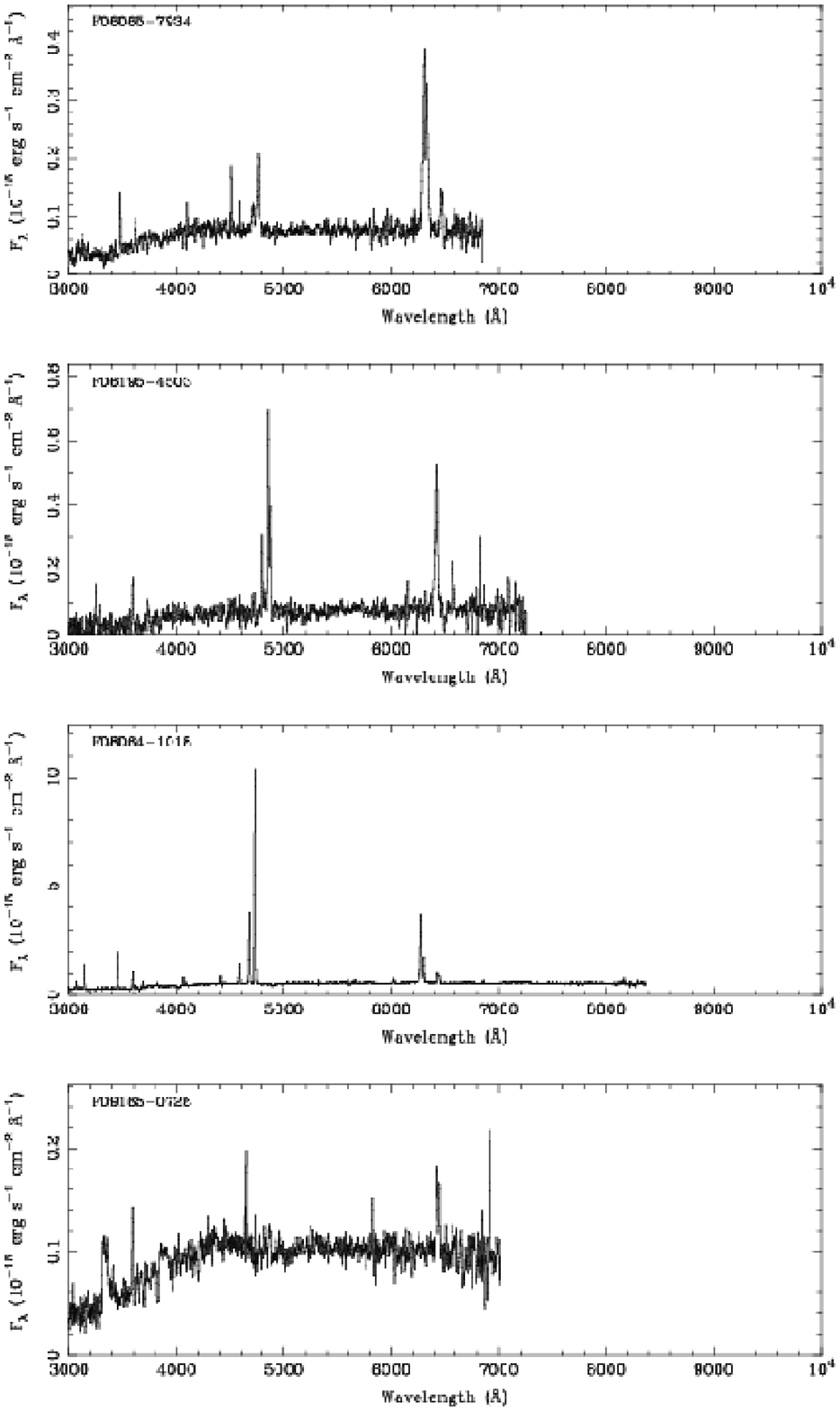}
\caption{\small \protect\emph{Continued.}}
\epsscale{1.0}
\end{figure}

\begin{figure}[H]
\addtocounter{figure}{-1}
\epsscale{0.7}
\plotone{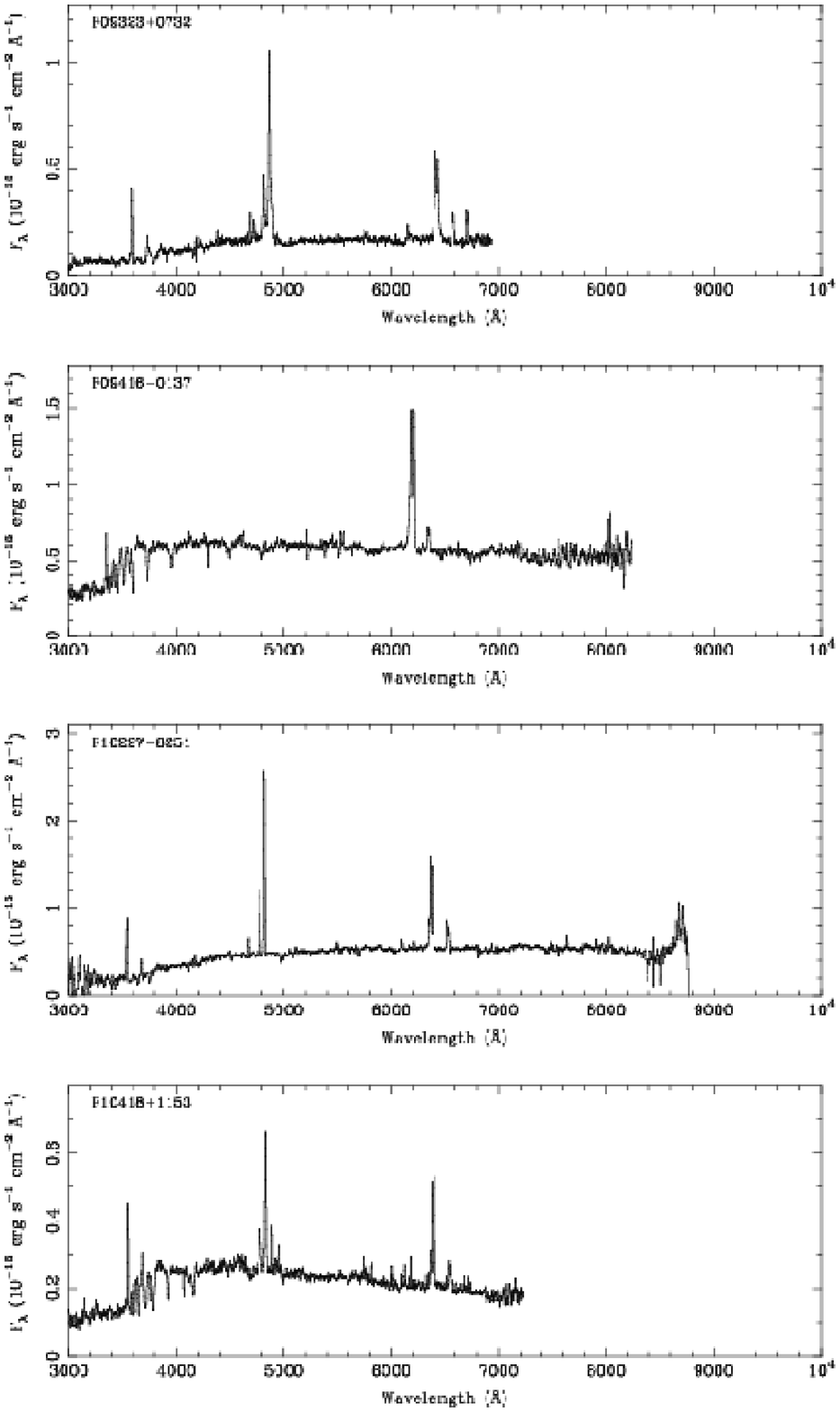}
\caption{\small \protect\emph{Continued.}}
\epsscale{1.0}
\end{figure}

\begin{figure}[H]
\addtocounter{figure}{-1}
\epsscale{0.7}
\plotone{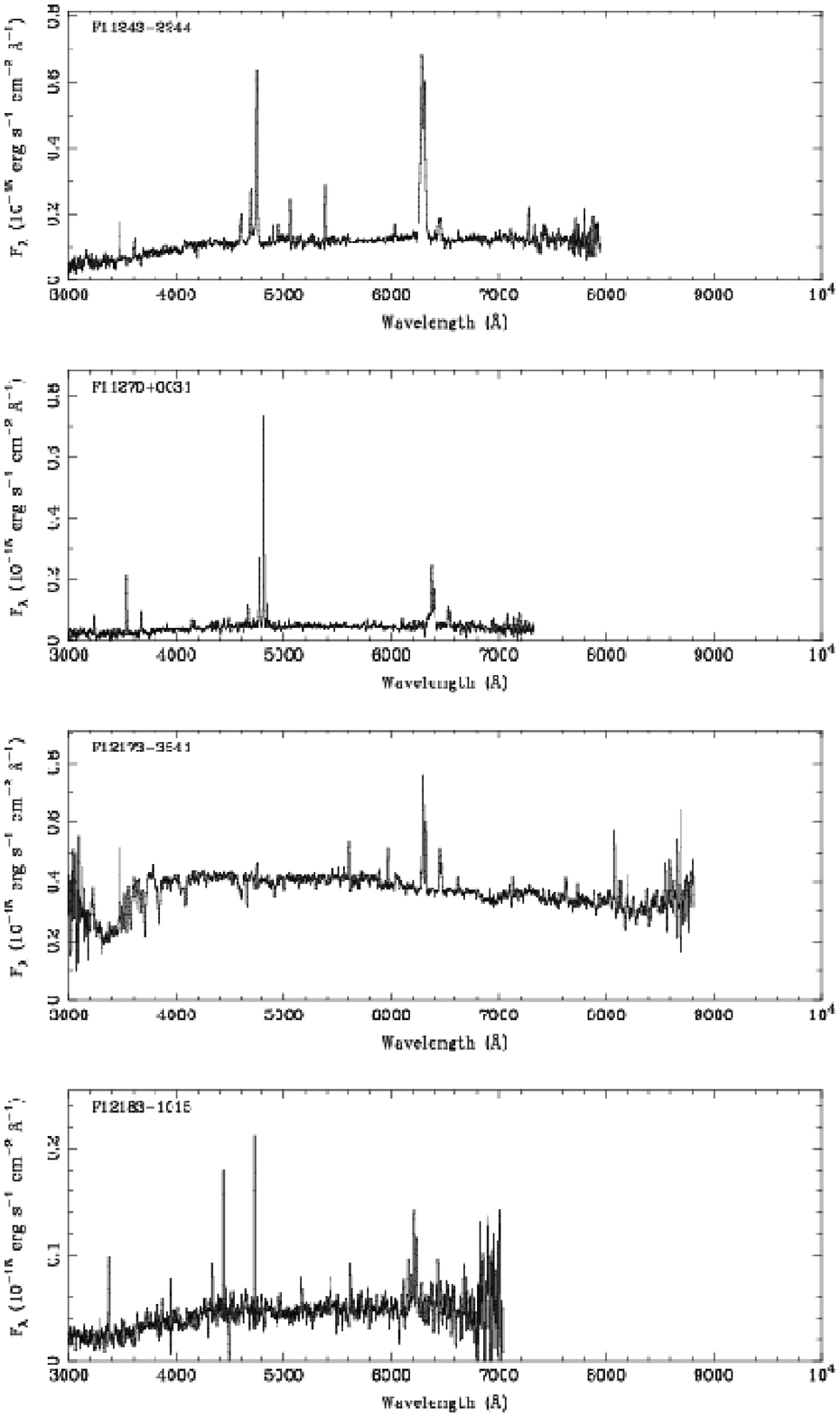}
\caption{\small \protect\emph{Continued.}}
\epsscale{1.0}
\end{figure}

\begin{figure}[H]
\addtocounter{figure}{-1}
\epsscale{0.7}
\plotone{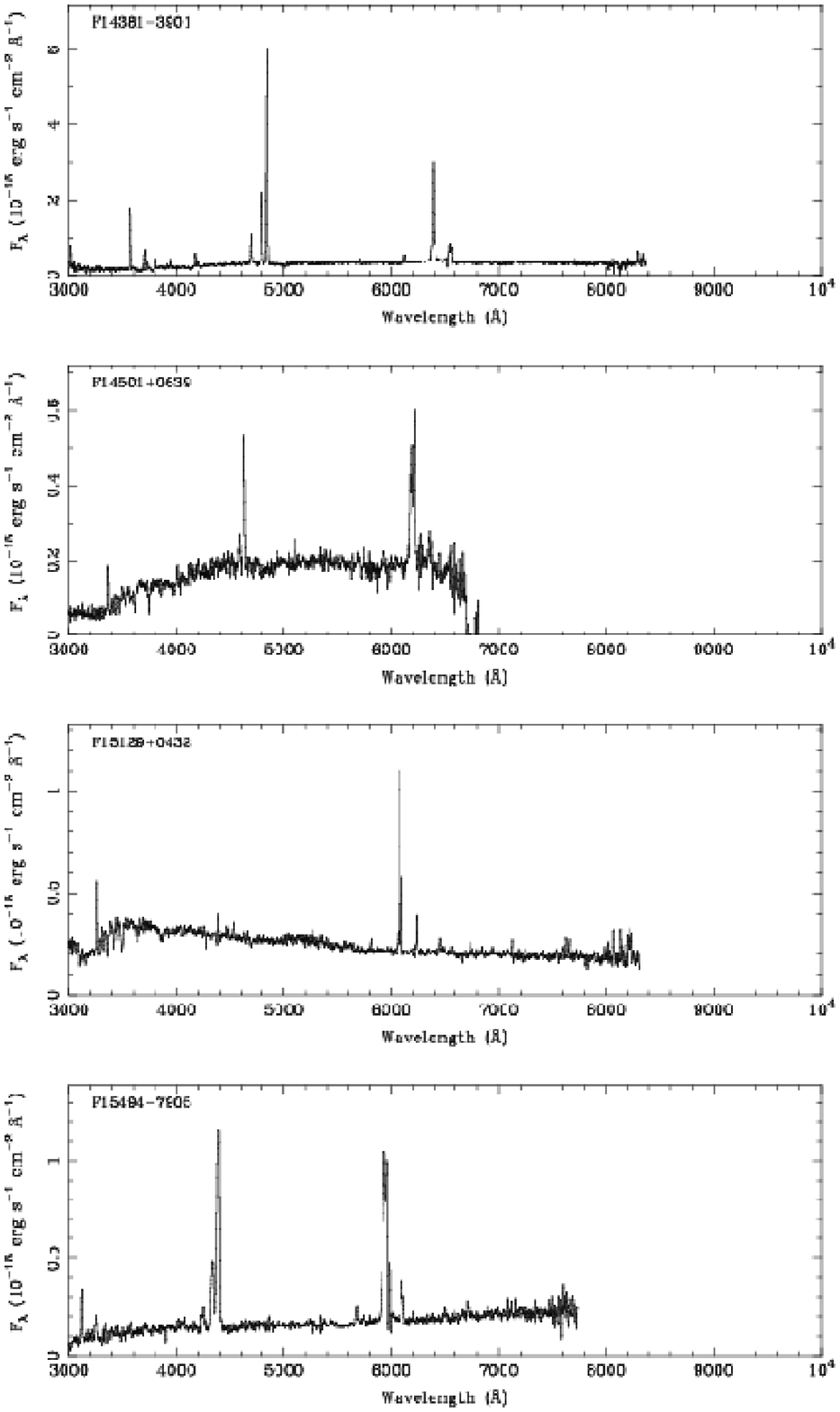}
\caption{\small \protect\emph{Continued.}}
\epsscale{1.0}
\end{figure}

\clearpage

\begin{figure}[H]
\addtocounter{figure}{-1}
\epsscale{0.7}
\plotone{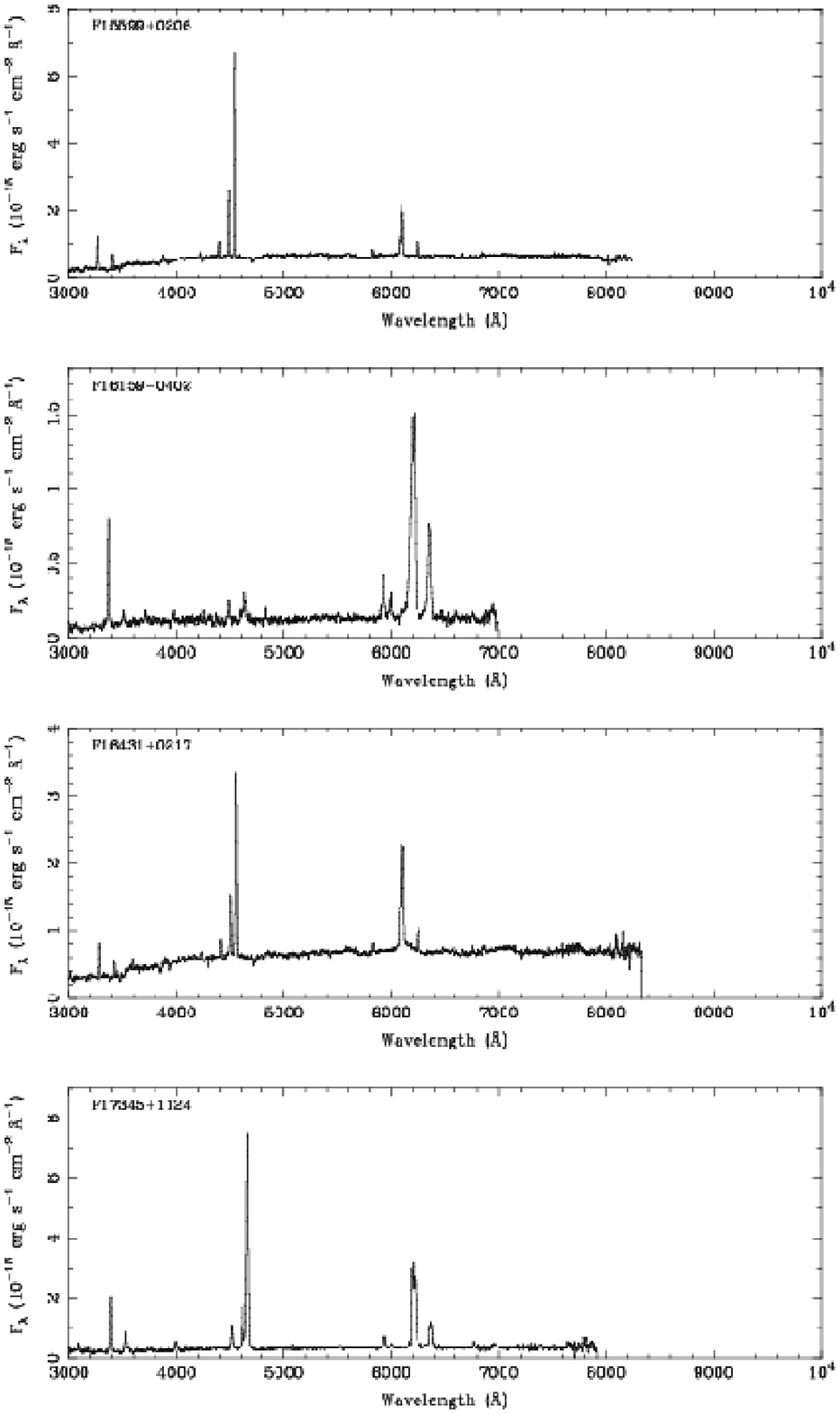}
\caption{\small \protect\emph{Continued.}}
\epsscale{1.0}
\end{figure}

\begin{figure}[H]
\addtocounter{figure}{-1}
\epsscale{0.7}
\plotone{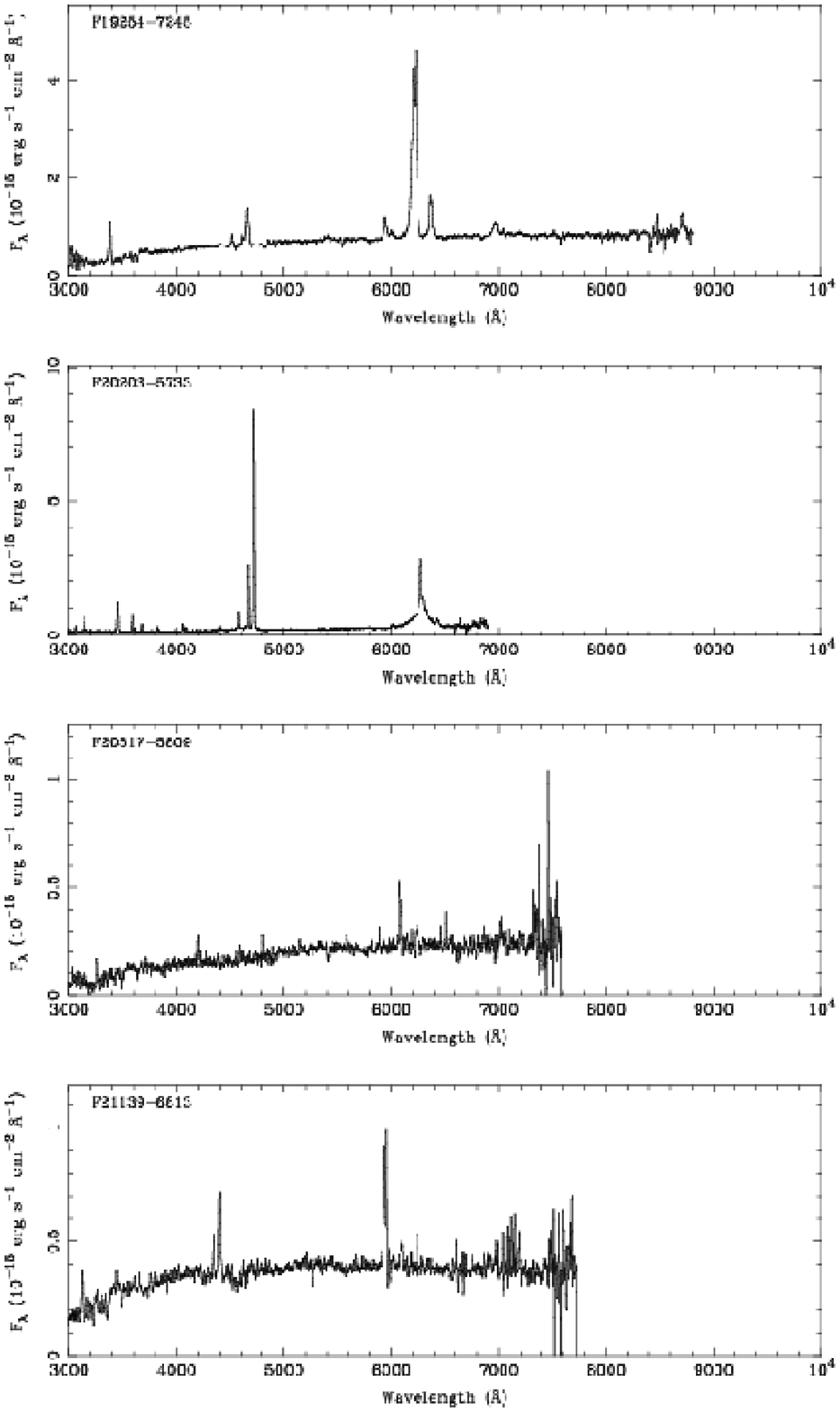}
\caption{\small \protect\emph{Continued.}}
\epsscale{1.0}
\end{figure}

\begin{figure}[H]
\addtocounter{figure}{-1}
\epsscale{0.7}
\plotone{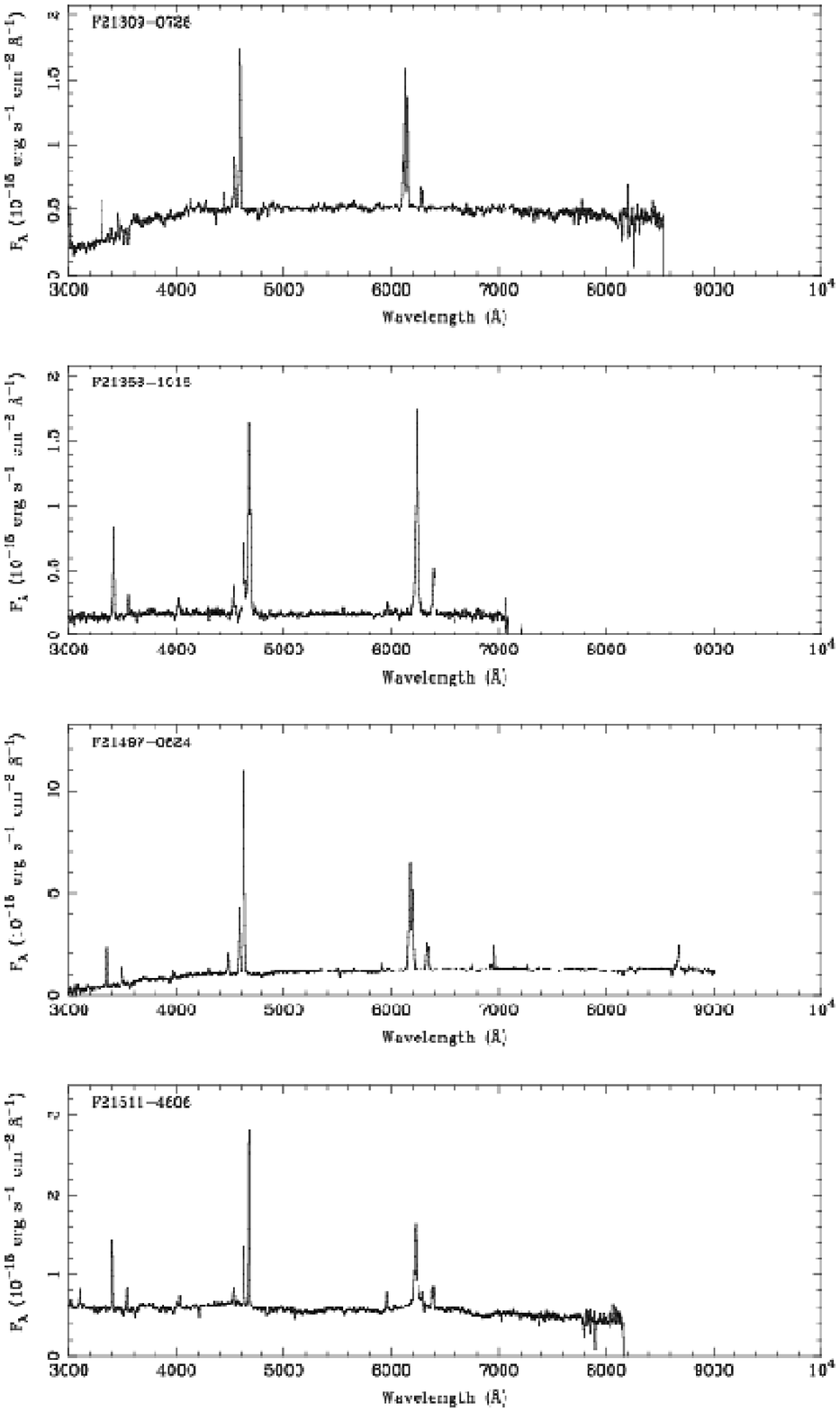}
\caption{\small \protect\emph{Continued.}}
\epsscale{1.0}
\end{figure}

\begin{figure}[H]
\addtocounter{figure}{-1}
\epsscale{0.7}
\plotone{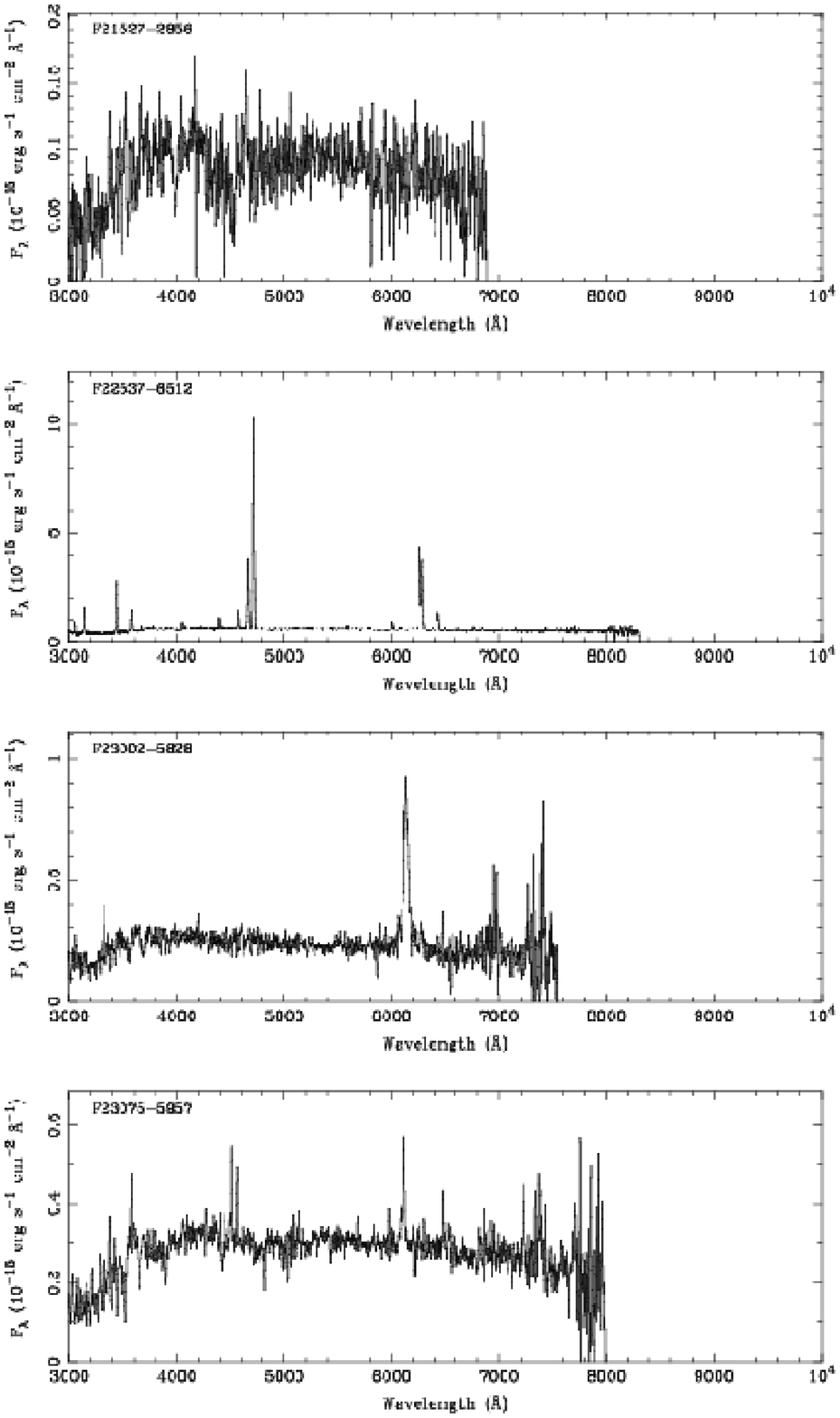}
\caption{\small \protect\emph{Continued.}}
\epsscale{1.0}
\end{figure}

\begin{figure}[H]
\addtocounter{figure}{-1}
\epsscale{0.7}
\plotone{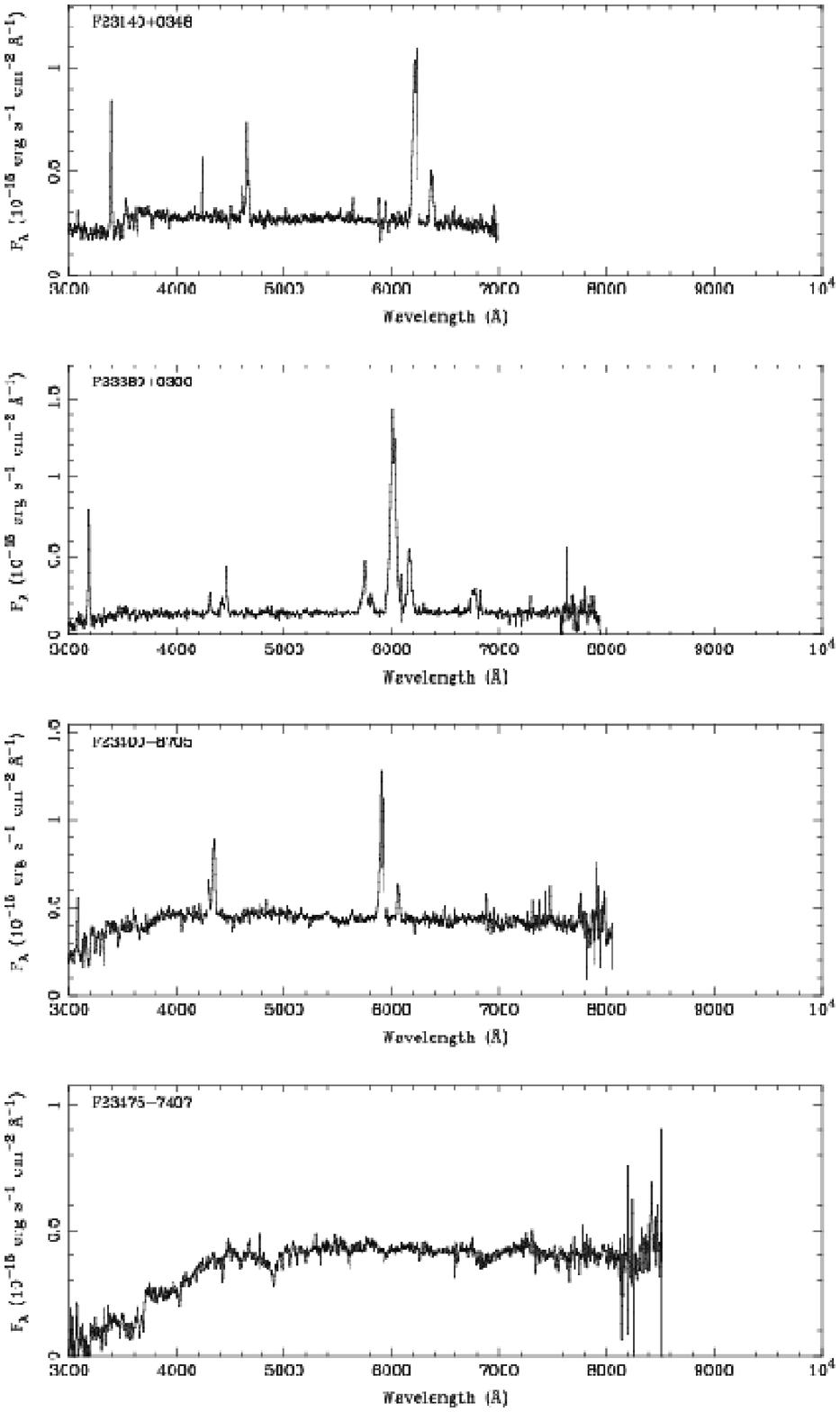}
\caption{\small \protect\emph{Continued.}}
\epsscale{1.0}
\end{figure}

\begin{figure}[H]
\addtocounter{figure}{-1}
\epsscale{0.7}
\plotone{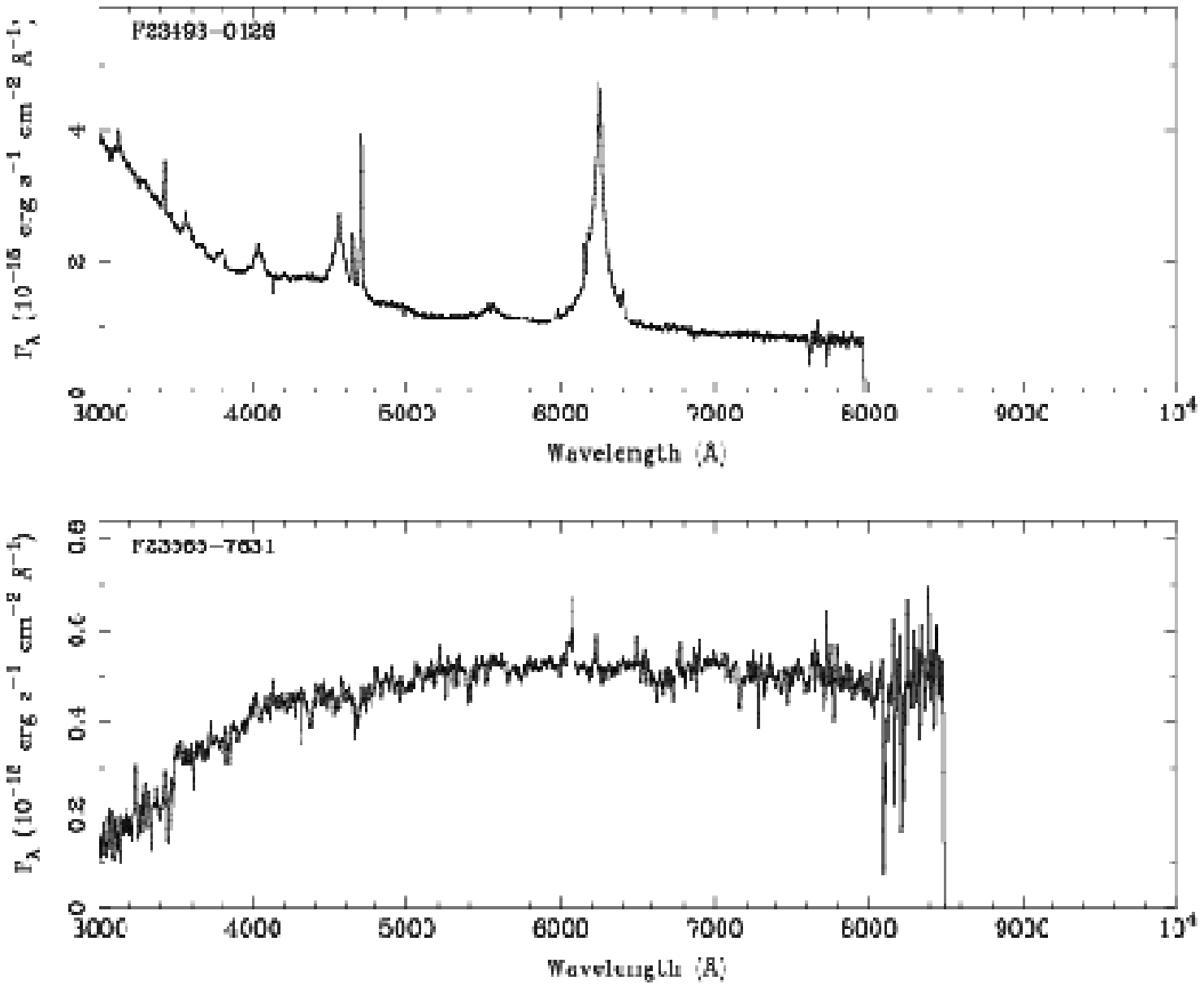}
\caption{\small \protect\emph{Continued.}}
\epsscale{1.0}
\end{figure}

\end{document}